\documentclass{aa}
\usepackage{graphicx}
\usepackage{amssymb}
\usepackage{amsmath}
\usepackage{natbib}

\bibpunct{(}{)}{;}{a}{}{,}

\def\eg{{\it e.g.\/}\rm,\ }

\def\void#1{{}}
\begin{document}
   \title{New spectroscopic confirmations of high-redshift galaxy
   clusters\thanks{Based on observations collected at the European
   Southern Observatory, Chile under the programs ESO65.O-0192(B) and
   ESO65.O-0189(A)}}

   \titlerunning{New spectroscopic confirmations of high-z galaxy
   clusters} 

   \author{L.F. Olsen\inst{1,2} \and E. Zucca\inst{3} \and
   S. Bardelli\inst{3} \and C. Benoist\inst{1} \and L. da
   Costa\inst{4,5} \and H.E. J{\o}rgensen\inst{2} \and
	   A. Biviano\inst{6}
	  \and
	   M. Ramella\inst{6}
	   }

   \offprints{L.F. Olsen, lisbeth@astro.ku.dk} 
\institute{
Observatoire de la C\^{o}te d'Azur, Laboratoire Cassiop\'ee, BP 4229, 06304 Nice Cedex 4, France 
\and Niels Bohr Institute, Copenhagen University, Juliane Maries Vej 30, 2100 Copenhagen, Denmark 
\and INAF - Osservatorio Astronomico di Bologna, via Ranzani 1, 40127 Bologna, Italy 
\and Observatorio Nacional, Rua General Jose Cristinino 77, CEP 20921-400, Rio de Janeiro JR, Brazil
\and European Southern Observatory, Karl-Schwartzschild-Str. 2, 85748 Garching b. M\"{u}nchen, Germany
   \and INAF - Osservatorio Astronomico di Trieste, Via G.B. Tiepolo 11, 34131 Trieste, Italy 
}

   \date{Received .....; accepted .....}

   \abstract {We present new spectroscopic data in the field of five
   high-redshift ($z\geq0.6$) candidate galaxy clusters, drawn from
   the EIS Cluster Candidate Catalog. A total of 327 spectra were
   obtained using FORS1 at the VLT, out of which 266 are galaxies with
   secure redshifts. In this paper, we use these data for confirming
   the existence of overdensities in redshift space at the approximate
   same location as the matched-filter detections in the projected
   distribution of galaxies from the EIS $I$-band imaging survey. The
   spectroscopic redshifts, associated to these overdensities, are
   consistent but, in general, somewhat lower than those predicted by
   the matched-filter technique. Combining the systems presented here
   with those analyzed earlier, we have spectroscopically confirmed a
   total of nine overdensities in the redshift range $0.6 < z < 1.3$,
   providing an important first step in building an
   optically-selected, high-redshift sample for more detailed studies,
   complementing those based on the few available X-ray selected
   systems.

	 }

   \maketitle
%

\section{Introduction}

Clusters of galaxies are large bound systems that evolve from
large-scale fluctuations, making their existence at large redshifts an
important constraint on cosmological models. They are also ideal sites
to study galaxy evolution, once systems at different redshifts are
available.  Combined, these characteristics have stimulated systematic
searches for large, statistical samples of galaxy clusters at large
redshifts ($z\gtrsim0.5$). With few exceptions most of the systems
identified, especially those at very large redshifts, have been
serendipitous discoveries in deep X-ray exposures. While this work
has provided confirmation for the existence of these systems, only a
handful of clusters have been identified. 

A more promising alternative is to use moderately deep, optical or
near-infrared surveys to search for concentrations in the projected
galaxy distribution as originally carried out by \cite{postman96} and
later used by \cite{olsen99a,olsen99b} and \cite{scodeggio99}, among
others. While finding cluster candidates using these single-passband
imaging surveys is much easier and yields much larger samples than
using the available X-ray data, the task of confirming that the systems
correspond to true density enhancements in redshift space and
going even further to bound systems is much harder.

Over the past few years our group has been engaged in an effort to
study galaxy systems at different redshift ranges with the aim of
confirming and if possible determining the nature of the EIS cluster
candidates \citep{olsen99a,olsen99b,scodeggio99}. Previous work have
included \cite{olsen03,olsen05} for systems at low redshift
($z\lesssim0.4$), and \cite{ramella00} for systems at intermediate
redshifts ($0.5\lesssim z\lesssim0.7$). For candidates with estimated
redshifts larger than $z \gtrsim0.6$ we have carried out observations
with FORS1 and FORS2 mounted at the VLT.  Preliminary results were
presented by \cite{benoist02} who showed strong evidence for a system
at $z=1.3$. The present paper extends these earlier results by
presenting the result of VLT spectroscopic observations of 5
additional fields.

In Sect.~\ref{sec:sample} we discuss how the candidate clusters were
selected from the original catalog and how available photometric data
were used to select individual galaxy targets, aiming at improving the
efficiency of the observations by eliminating foreground and
background objects.  In Sect.~\ref{sec:obs_data} the observations and
data reduction are summarized. In Sect.~\ref{sec:results} the results
of the spectroscopic observations are presented for each of the fields
considered.  In Sect.~\ref{sec:discussion} these results are
combined with those of \cite{benoist02} to draw conclusions
regarding the efficiency of the matched-filter technique applied to
moderately deep $I$-band survey data in building a statistical sample
of high-redshift galaxy clusters.  Finally, in Sect.~\ref{sec:summary},
the main results of the present paper are summarized.


\section{Sample Selection}
\label{sec:sample}

Cluster candidates were drawn from the sample of EIS cluster
candidates compiled by \cite{olsen99a,olsen99b} and \cite{scodeggio99}
applying the matched-filter technique to the EIS-WIDE $I$-band imaging
survey covering 17~square degrees. The eight target clusters, of which
three were discussed by \cite{benoist02}, were selected based on their
identification as likely clusters in a color slice analysis
\citep{olsen00}. This analysis was based on the optical survey data
combined with infrared follow-up imaging. Based on the fact that most
clusters exhibit a red sequence of early-type galaxies
\cite[e.g. ][]{gladders98, stanford98} we searched for concentrations
of galaxies with similar color by separating the galaxies in slices of
color and identifying peaks in the density distribution for each
color.  The analysis was carried out separately for the $I-K_s$ and
$J-K_s$ colors.  The systems selected for follow-up spectroscopy all
appeared to have significant overdensities in both $I-K_s$ and
$J-K_s$.  In Table~\ref{tab:cl_targets} we present the detection
information, both for the matched filter and the color slicing,
regarding the five candidates discussed in this work. The table gives:
in Col.~1 the field name; in Cols.~2 and 3 the nominal position of the
cluster candidate in J2000; in Col.~4 the redshift estimated by the
matched-filter algorithm; in Col.~5 the $\Lambda_{cl}$-richness,
which measures the equivalent number of $L^*$ galaxies and in Col.~6
the Abell like richness giving the number of galaxies in the magnitude
interval $[m_3;m_3+2]$, where $m_3$ is the third brightest galaxy. For
both richnesses more details can be found in \cite{olsen99a}; in
Cols.~7 and 8 the $I-K_s$ and $J-K_s$-colors obtained from the
color-slicing analysis.  When comparing the computed colors of the
cluster galaxies to those expected for a passively evolving elliptical
galaxy we find that in general, the $I-K_s$-colors are bluer, while
the $J-K_s$-colors were found to be roughly consistent with these
expectations. This may be caused by a poor calibration of the IR data
used for the preliminary analysis.

\begin{table*}
\caption{Basic properties for the targeted candidate clusters. The
parameters are described in the text and more details can be found in
\cite{olsen99a}.}
\label{tab:cl_targets}
\center
\begin{tabular}{lrrrrrrr}
\hline\hline
EIS Cluster & $\alpha$ (J2000) & $\delta$ (J2000) & $z_{MF}$ & $\Lambda_{cl}$ &
$N_R$ & $I-K_s$ & $J-K_s$\\
\hline
EISJ0046-2951 & $00:46:07.4$ & $-29:51:44.5$ & $0.9$ & $157.0$ & $2$ & 2.75 & 1.75\\
EISJ0048-2942 & $00:48:31.6$ & $-29:42:52.1$ & $0.6$ & $55.6$ & $13$ & 2.75 & 1.75\\
EISJ0050-2941 & $00:50:04.4$ & $-29:41:35.6$ & $1.0$ & $175.3$ & $62$ & 3.50 &
1.60\\
EISJ2236-4017 & $22:36:18.0$ & $-40:17:54.9$ & $0.6$ & $107.8$ & $47$& 2.90 & 1.45\\
EISJ2249-3958 & $22:49:33.0$ & $-39:58:10.1$ & $0.9$ & $123.6$ & $29$ & 2.75&1.90\\
\hline\hline
\end{tabular}
\end{table*}

The selection of target galaxies in each field was based on a
combination of data from as many bands as available.  For the
candidates EISJ0046-2951, EISJ0048-2942, EISJ0050-2941 we derived
photometric redshifts based on $BVIJK_s$ imaging. The limiting
magnitude used for this work was $I=22.5$ to avoid large errors in the
derived photometric redshifts. The primary targets were selected among
galaxies with $z_{phot}\geq0.5$ ($\sim50\%$ of the target
galaxies). Remaining slits were filled with arbitrarily chosen
objects. For the targets selected to have $z_{phot}\geq0.5$ $\sim70\%$
proved to have a spectroscopic redshift $z_{spec}\geq0.5$. A more
thorough discussion of the photometric redshifts is the topic of a
forthcoming paper.

For the clusters EISJ2236-4017 and EIS2249-3958 only $IJK_s$ imaging
data were available at the time of the spectroscopic observations. For
these clusters we selected galaxies based on their optical-infrared
colors to match those of elliptical galaxies at $z\geq0.5$. Remaining
slits were filled with arbitrarily chosen galaxies.

\section{Observations and data reduction}
\label{sec:obs_data}

The observations were carried out in the nights September 21-25, 2000,
using FORS1 mounted at the VLT-ANTU telescope. We used the
multi-object spectroscopy (MOS) mode, in which FORS1 provides 21 slits
with a length of 20 and 22~arcsecs (see the FORS Manual for
details). The length of the slits is much longer than necessary for
each galaxy, thus as often as possible we tried to fit two galaxies in
the slits. In practice, however, this was rarely possible.  The MOS
masks were positioned using the FIMS-software developed for this
purpose. We used the $I$-band images from the ESO Imaging Survey
\citep{nonino99, benoist99} to determine the positions of the slits.
We used grism 150I+17 with the order separation filter OG590 covering
the wavelength range 6000-11000{\AA}. The dispersion of 230{\AA}/mm
corresponding to 5.52{\AA}/pixel gave a spectral resolution of 280 or
about 29{\AA} for a slit width of 1.4~arcsecs.  The exposure times for
each mask were either 60~min or 120~min, depending on the $I$-band
magnitude of the target galaxies. We split the exposures into four or
eight 15~min exposures. Calibration frames (flatfields and calibration
arcs) were obtained during daytime.

We reduced the spectra using IRAF-tasks written for this purpose based
on the APALL task. Details on the reduction procedure and measurement
of redshift are available in J{\o}rgensen et al. (2005, in
preparation).  Here, it suffices to say that the individual science
exposures were combined and a flatfield correction applied before the
wavelength calibration.  Redshifts were computed by cross-correlating
the extracted one-dimensional spectra against template spectra, taken
from \cite{kinney96}, properly shifted to a redshift close to that
expected for the galaxy being measured as estimated from
features in the galaxy spectrum. We apply this procedure iteratively,
and also for different galaxy spectra. Before a redshift was accepted
it was compared with the presence of corresponding spectral
features. The redshifts derived in this way are listed in
Tables~\ref{tab:EIS0046-2951} - \ref{tab:EIS2249-3958}. We estimate an
accuracy of the individual redshifts of $\delta
z=0.0004\sim130\mathrm{km/s}$. In these tables a value of $8.8888$
indicates that the spectrum revealed a stellar object.


\section{Results}
\label{sec:results}

\begin{figure*}
\center
\resizebox{0.23\textwidth}{!}{\includegraphics{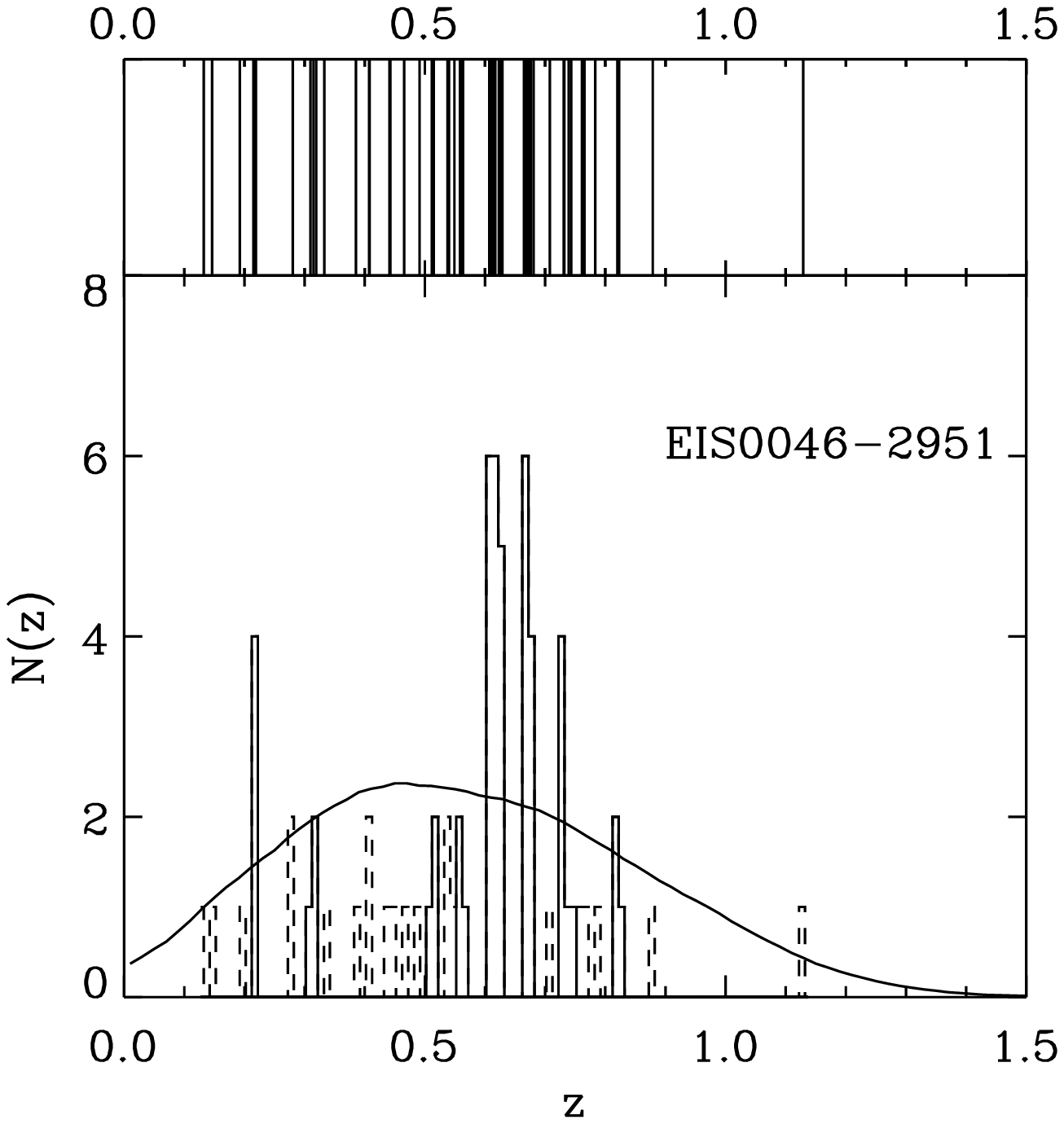}}
\resizebox{0.23\textwidth}{!}{\includegraphics{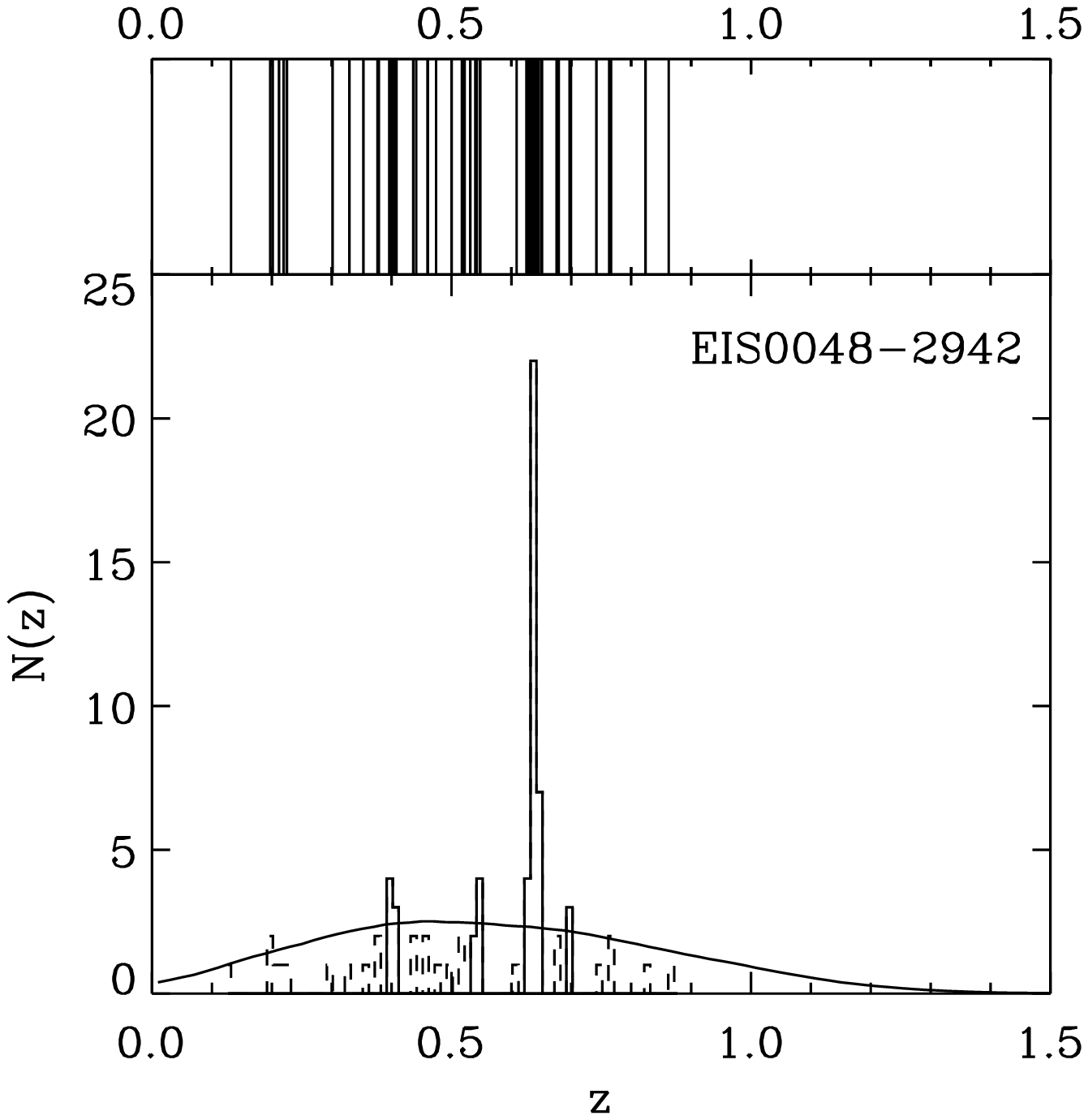}}
\resizebox{0.23\textwidth}{!}{\includegraphics{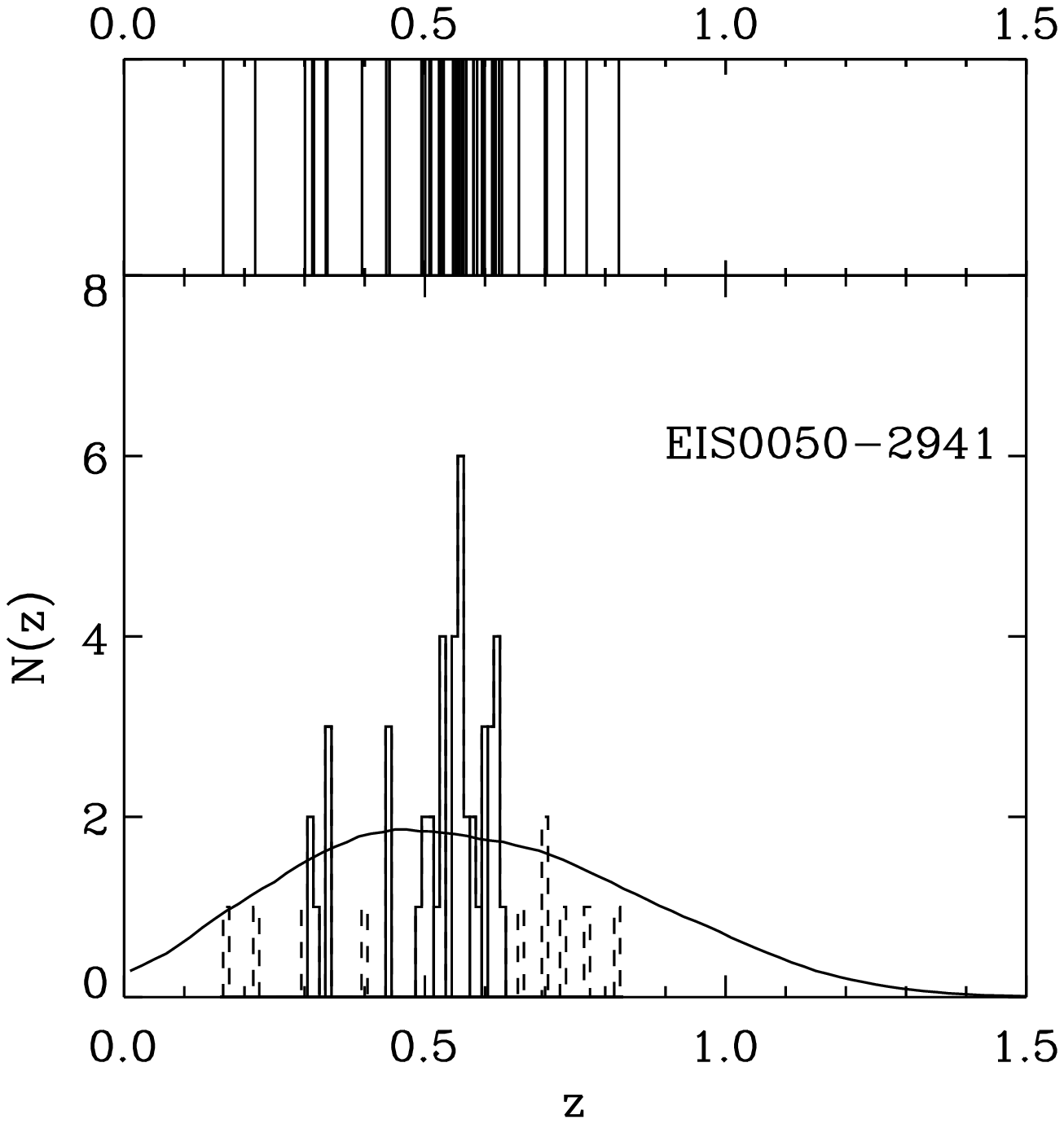}}

\resizebox{0.23\textwidth}{!}{\includegraphics{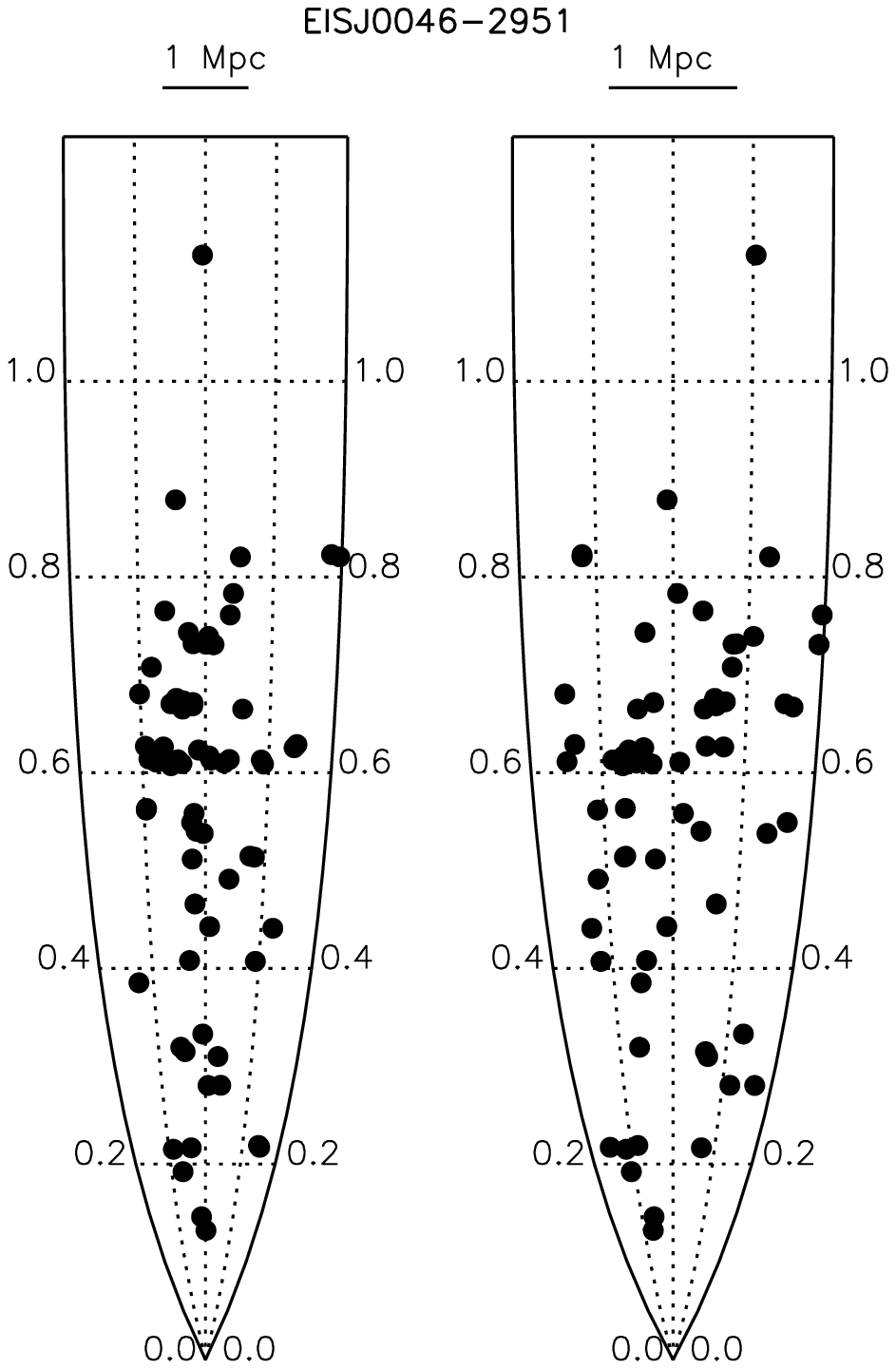}}
\resizebox{0.23\textwidth}{!}{\includegraphics{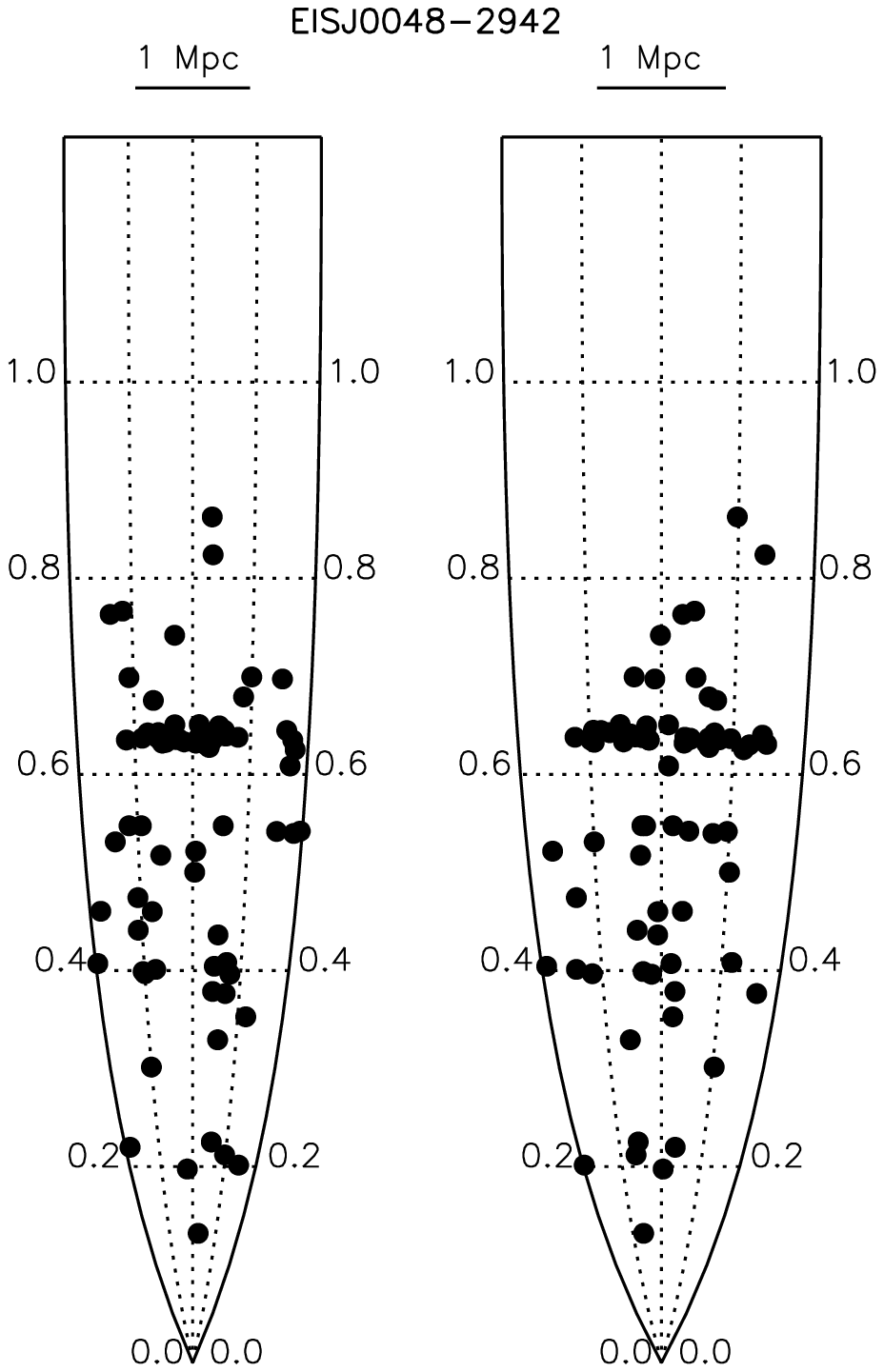}}
\resizebox{0.23\textwidth}{!}{\includegraphics{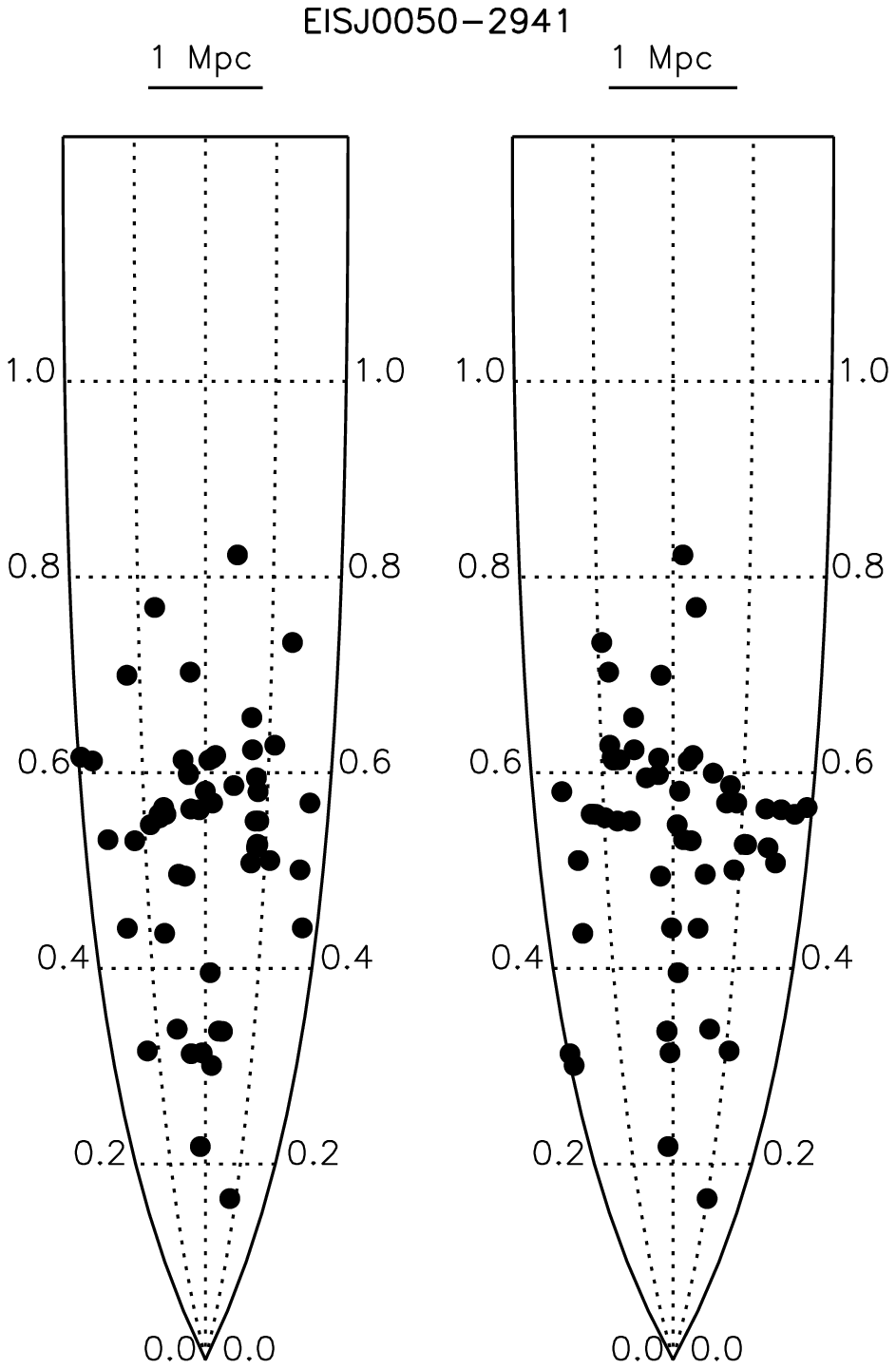}}

\resizebox{0.23\textwidth}{!}{\includegraphics{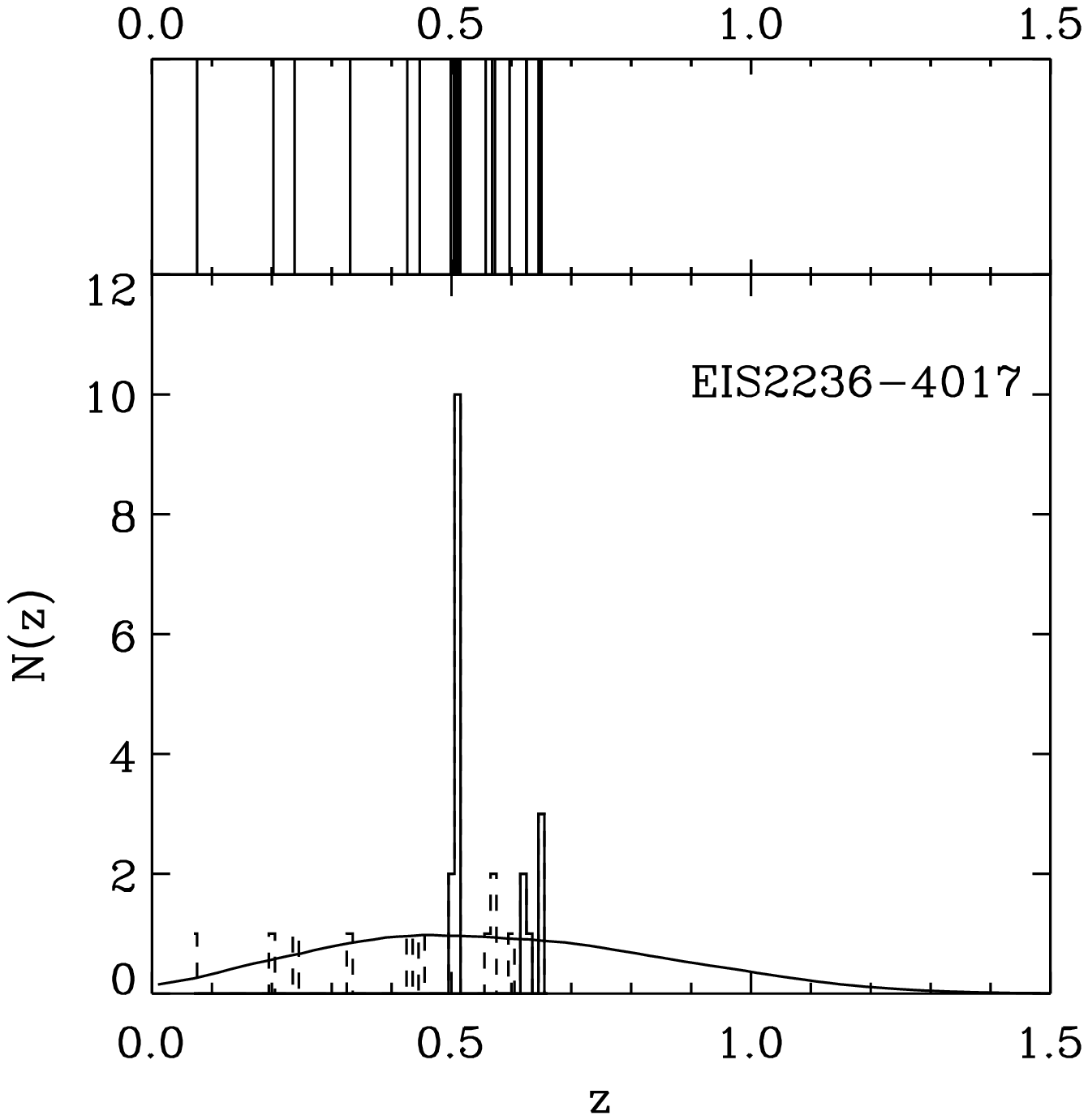}}
\resizebox{0.23\textwidth}{!}{\includegraphics{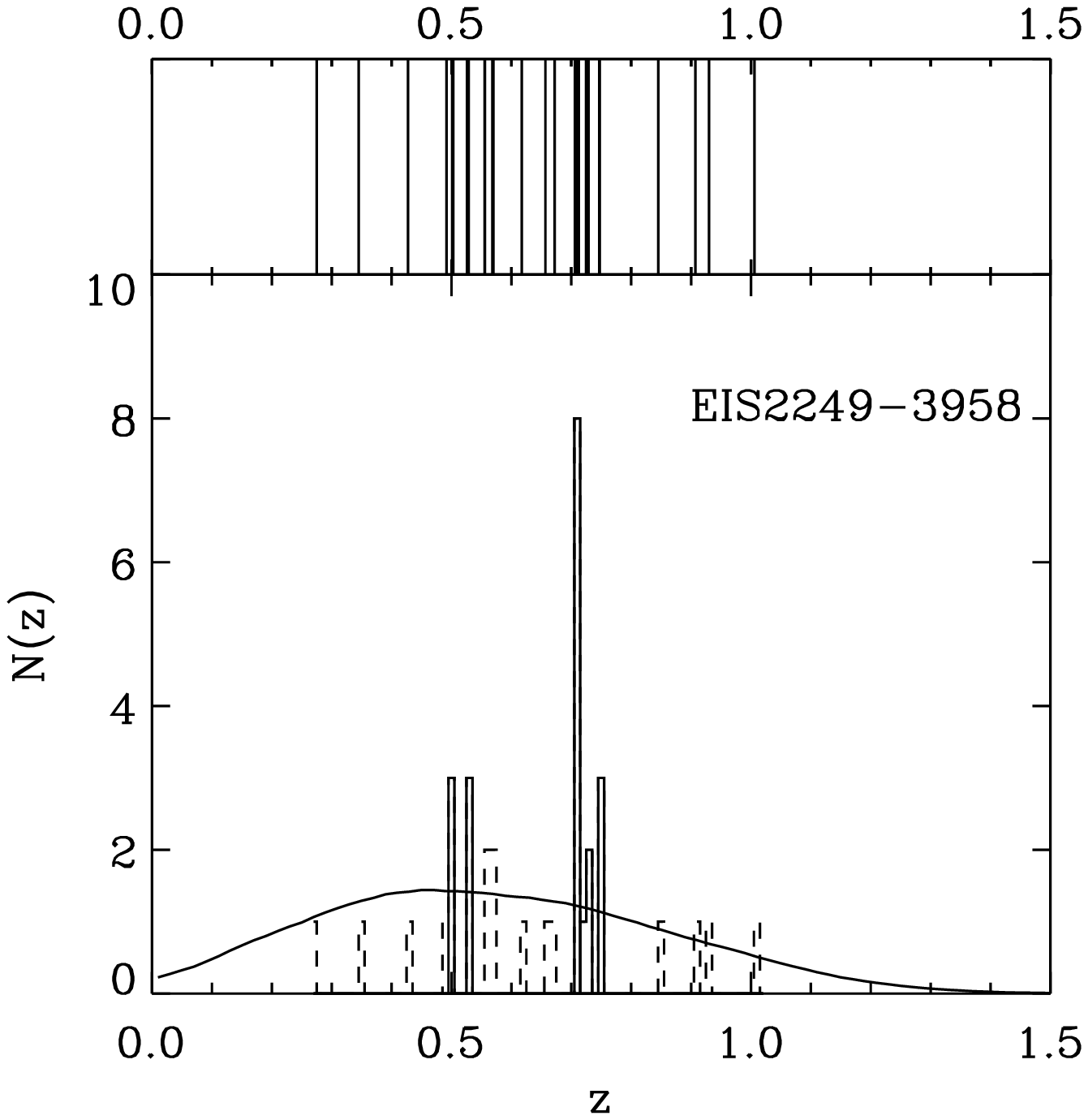}}

\resizebox{0.23\textwidth}{!}{\includegraphics{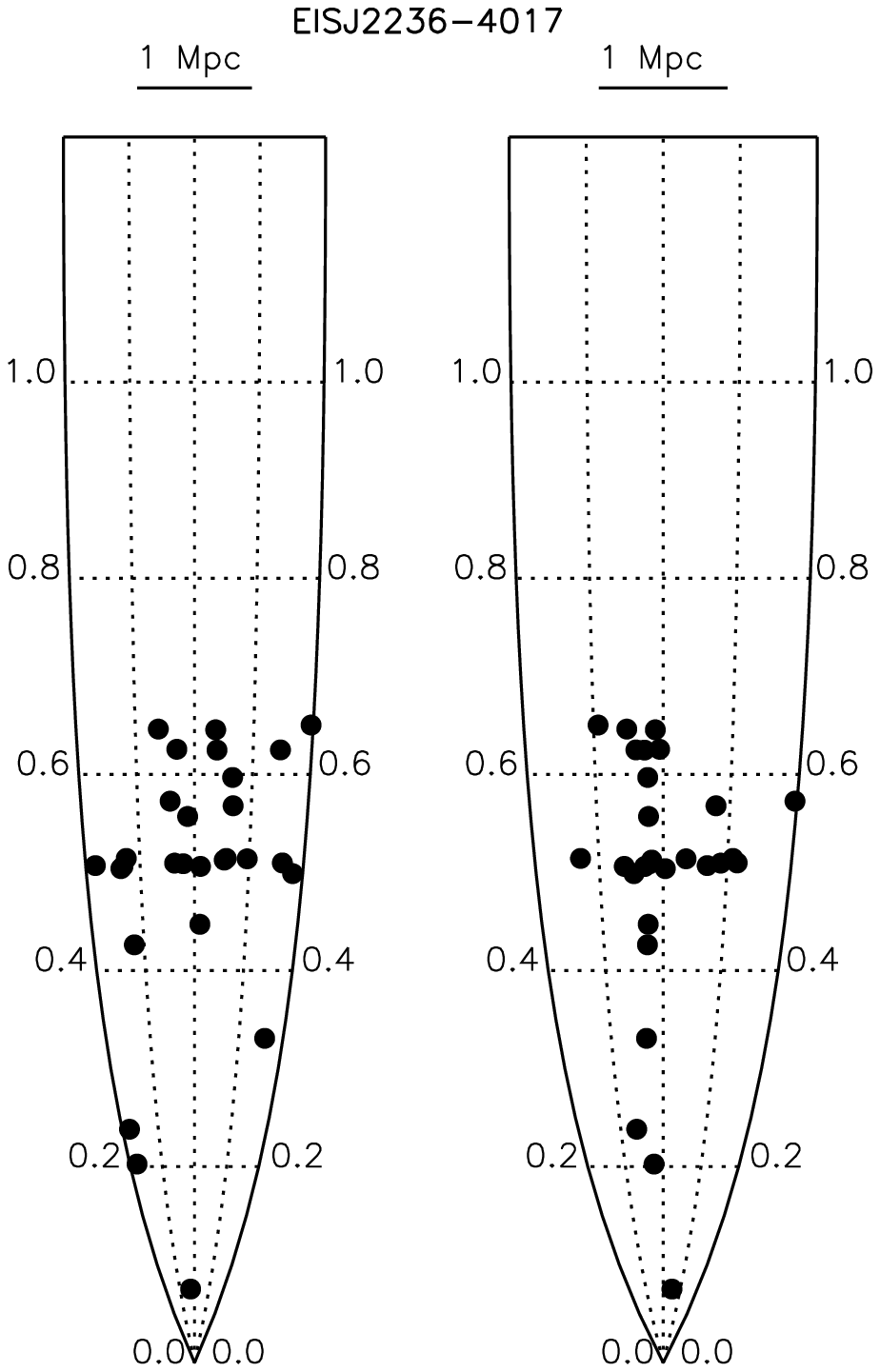}}
\resizebox{0.23\textwidth}{!}{\includegraphics{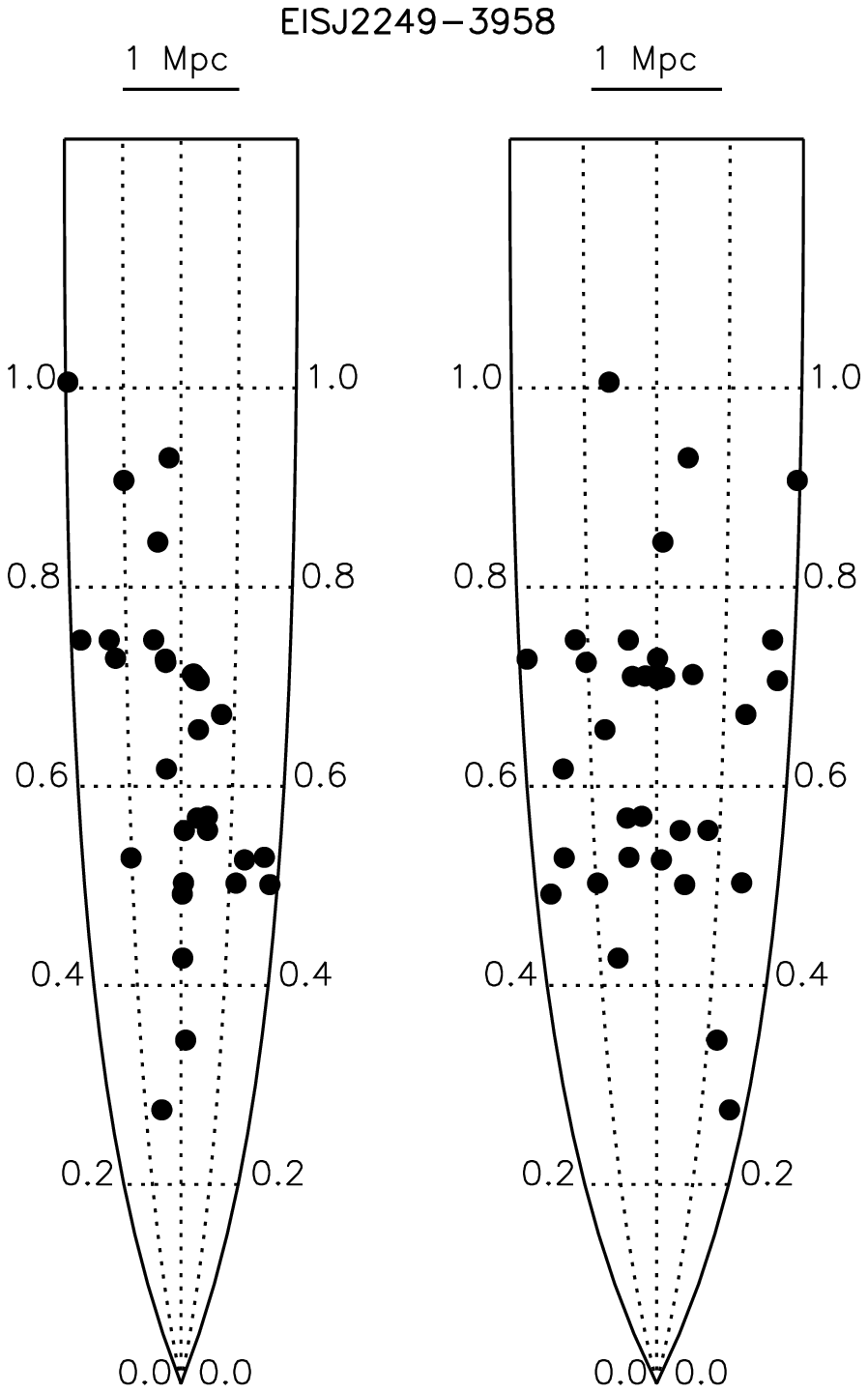}}

\caption{For each cluster the obtained redshifts (upper panels) and
the redshifts as function of right ascension and declination (lower
panels, left and right) are shown. In the upper panel the upper part
gives a bar diagram of the redshifts, while the lower part gives the
redshift distribution (dashed line) with the identified groups (solid
line). In the lower panels the left cone shows the redshifts as
function of right ascension and the right one the redshifts as function
of declination. The diagrams correspond to the complete coverage of
each field. The bar in the top of each cone gives the scale of
$1h_{75}^{-1}\mathrm{Mpc}$ The shape of the cones translates the
evolution of scale with redshift.}
\label{fig:zdistributions}
\end{figure*}

We have secured a total of 266 new galaxy redshifts. The distribution
of the redshifts of each field is shown in the upper panels of
Fig.~\ref{fig:zdistributions}. These panels give in their upper parts
a bar diagram of all the measured redshifts and in their lower parts
the distribution of redshifts (dashed line) with the solid line
indicating identified groups as discussed below. The figure also shows
for each field the redshifts versus right ascension and declination
respectively (lower panels, left and right).

As in previous papers of this series, we use the ``gap''-technique
originally proposed by \cite{katgert96} for identifying groups in
redshift space. We adopt a gap-size of 0.005(1+z) to separate
individual groups. This separation corresponds to a restframe velocity
of 1500~km/s.  In addition, only systems with a group redshift $z>0.4$
and with at least 3 galaxies are considered for further analysis. This
lower limit on the redshift corresponds to an offset from the original
matched-filter estimate of $\Delta z=0.2$ which should be sufficient
to include all confirmations.

To assess the significance of each group we have used simulated data
sets based on the expected redshift distribution for a uniform
distribution of galaxies with a given luminosity function (LF). The
redshift distribution is built for the same limiting magnitude as was
used for the target selection ($I=22.5$). It was confirmed that this
approach leads to a redshift distribution which is consistent with
that measured by the Canada-France Redshift Survey \citep{lilly95}
when the same limiting magnitude is adopted \cite[for further details
see ][]{benoist02}. We determine the significance of the detected
groups from the probability of finding a group as rich or richer at
the same redshift. To do this we draw 1000 sets of galaxies from the
redshift distribution constructed above, with the size given by the
number of redshifts measured in each of the cluster fields. We select
only galaxies with $z\geq0.4$ to mimic the color pre-selection of our
targeted galaxies. For each set we run our group finding method to
obtain the frequency of groups as rich or richer than and at the same
redshift (within $\Delta z \leq 0.05$) as the group detected in the
spectroscopic data.  This constraint in redshift is necessary because,
due to the shape of the selection function, the frequency of groups
with a certain number of members varies with redshift. The
significance of the group is defined to be $1-f$, where $f$ is the
redshift dependent frequency.

Applying the gap-technique to the redshift distributions shown in
Fig.~\ref{fig:zdistributions}, and adopting the same criteria used in
previous papers to consider only density enhancements with a
significance $\geq 99\%$, we identify 8 groups. Their properties are
summarized in Table~\ref{tab:groups} which gives: in Col.~1 the name
of the EIS cluster field; in Cols.~2 and 3 the J2000 right ascension
and declination; in Col.~4 the number of member galaxies; in Col.~5
the redshift of the group; in Col.~6 its biweight estimated restframe
velocity dispersion corrected for our measurement accuracy with 68\%
bootstrap errors; and finally, in Col.~7 the significance, as defined
above. The positions given in the table are mean values computed using
the spectroscopically confirmed member galaxies.  Below the detections
for each individual cluster field are briefly discussed using the
available color information to provide additional information.

\begin{table*}
\caption{The identified groups with significance $\sigma \geq 99\%$.}
\label{tab:groups}
\center
\begin{tabular}{lrrrrrr}
\hline\hline
Cluster & $\alpha$ (J2000) & $\delta$ (J2000) & \# galaxies & $z$ & $\sigma_v$
(km/s) & Significance\\
\hline
EIS0046-2951      & 00:46:04.2 & -29:49:27.6 &17&     0.614 &  $1400^{+210}_{-610}$&	99.9\smallskip\\
EIS0046-2951      & 00:46:07.7 & -29:51:04.9 &10&     0.671 &   $865_{-270}^{+120}$ &	99.7\smallskip\\
\hline
EIS0048-2942      & 00:48:35.4 & -29:41:52.4 & 7&     0.402 &  $1000_{-480}^{+100}$  &	99.4\smallskip\\
EIS0048-2942      & 00:48:33.4 & -29:42:28.9 &33&     0.637 &  $1080_{-210}^{+150}$  &	$>$99.9\smallskip\\
\hline
EIS0050-2941      & 00:50:06.2 & -29:40:35.3 &12&     0.558 &  $1375_{-270}^{+190}$  &	99.9\smallskip\\
EIS0050-2941	  & 00:50:03.0 & -29:40:18.1 & 8&     0.616 &  $970_{-620}^{+210}$  &	99.1\smallskip\\
\hline
EIS2236-4017      & 22:36:22.0 & -40:17:55.0 &12&     0.509 &  $900_{-260}^{+160}$ &	99.9\smallskip\\
\hline
EIS2249-3958	  & 22:49:32.1 & -39:58:02.9 & 8&     0.710 &   $380_{-140}^{+ 50}$ &	99.4\smallskip\\
\hline
\end{tabular}
\end{table*}

\subsection{EISJ0046-2951}

\begin{figure*}
\begin{center}
\resizebox{0.6\textwidth}{!}{\includegraphics{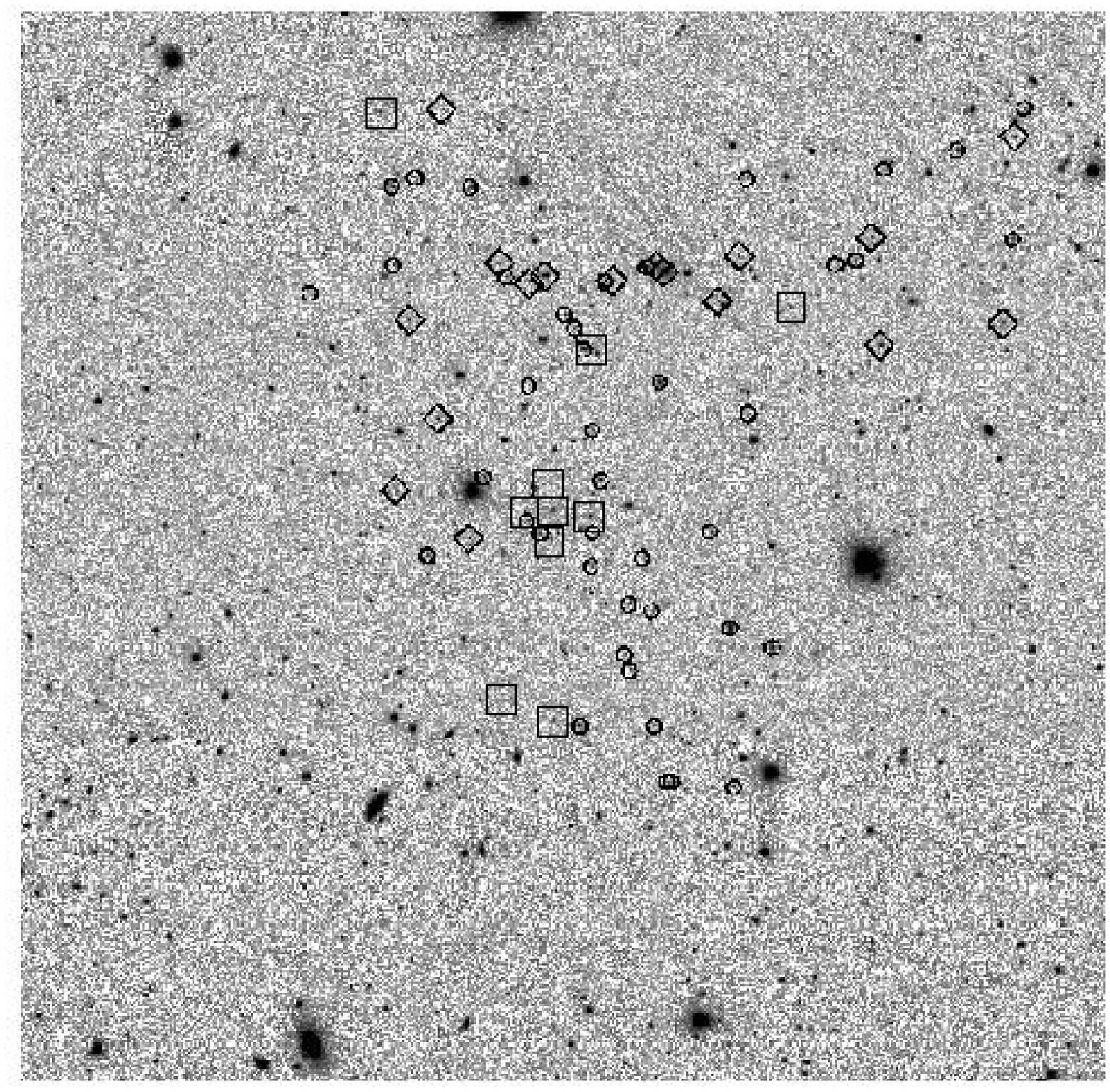}}
\caption{A 10$\times$10 arcmin cutout centered on the matched-filter
position of EISJ0046-2951. The circles
mark galaxies with redshifts outside significant groups. The diamonds
mark galaxies in the foreground group and the squares those in the
background group. North is up, east to the left.}
\label{fig:img0046-2951}
\end{center}
\end{figure*}

In this field 71 galaxy redshifts have been measured and are shown in
Fig.~\ref{fig:zdistributions}. We identify 6 groups with
at least 3 members at $z\geq0.4$, out of which only 2, one at
$z=0.614$ and the other at $z=0.671$ have significance $\geq99\%$ and
are, therefore, included in Table~\ref{tab:groups}. 

In order to decide which of these groups is the one most likely
associated with the matched-filter detection we examined both the cone
diagrams (Fig.~\ref{fig:zdistributions}) and the projected
distribution of the galaxies with redshifts as shown in
Fig.~\ref{fig:img0046-2951}. In the cone diagrams the foreground
system is more prominent than the background one, due to its larger
extent. From the projected distribution of the galaxies, it is clear
that the center of the foreground group deviates significantly from
that of the matched-filter algorithm, located at the center of the
field.  The background group, on the other hand, consists of 10
members out of which five form a compact system located very close to
the center derived by the matched filter and five more uniformly
distributed.  Furthermore, from the examination of the image one finds
that the five galaxies situated near the center are also among the
brightest galaxies in the central region. It should also be kept
in mind that the radial part of the matched filter gives a
significantly higher weight to galaxies close to the estimated
position, thus the galaxies closest to the originally estimated
position are the ones contributing the most, even when other systems
are found at almost the same redshift and in the same field. Combined
these arguments suggest that the background group at $z=0.671$
($\Delta z=z_{spec}-z_{MF}\sim - 0.23$) and with an estimated velocity
dispersion of $\sigma_v=865\mathrm{km/s}$ is the most likely to
correspond to the EIS cluster candidate identified by the
matched-filter algorithm, with the difference between the estimated
and measured redshifts being consistent with the expected
errors. Another possible explanation, could be that the matched-filter
detection at a larger redshift is correct but system galaxies have not
been observed or had no measured redshifts.  However, careful
inspection of the image show no evidence for any further clustering of
faint galaxies in the field.

\begin{figure*}
\begin{center}
\resizebox{0.6\textwidth}{!}{\includegraphics{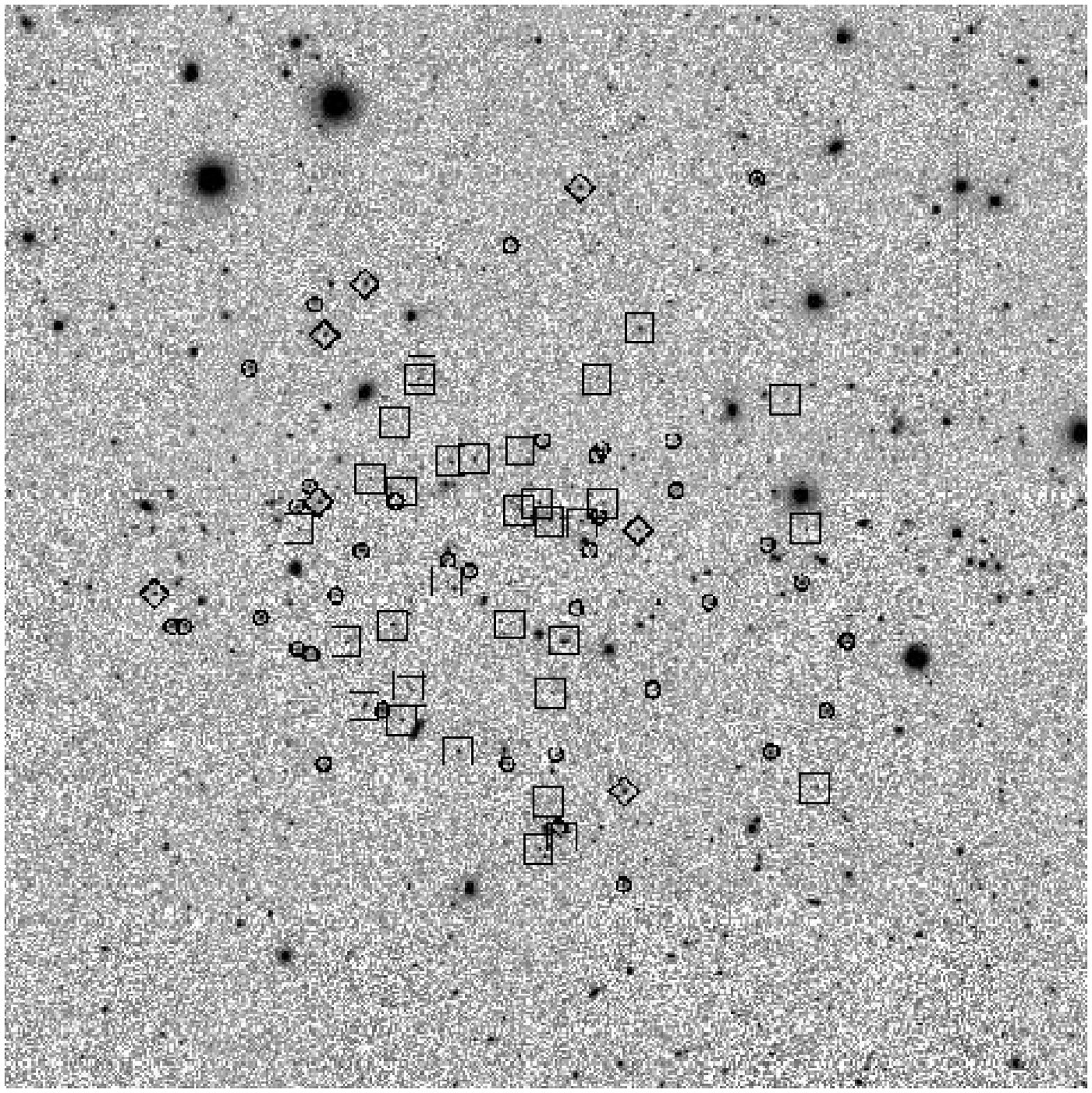}}
\caption{A 10$\times$10 arcmin cutout centered on the
matched-filter position of EISJ0048-2942. Symbols follow those in
Fig.~\protect\ref{fig:img0046-2951}. North is up, east is to the
left.}
\label{fig:img0048-2942}
\end{center}
\end{figure*}

\subsection{EISJ0048-2942 }

The distribution of the 77 redshifts measured in this field is
presented in Fig.~\ref{fig:zdistributions}. As it can be seen, the
distribution shows a distinct spike around $z\sim0.6$, in excellent
agreeement with the value estimated by the matched filter. The
gap-technique identifies 3 groups with at least 3 members in this
field, out of which two have significance $\geq99\%$ and are therefore
included in Table~\ref{tab:groups} - one with seven members at
$z=0.402$ and a more prominent background group with 33 members at
$z=0.637$.

The examination of the redshift distribution and the cone diagrams for
this cluster (shown in Fig.~\ref{fig:zdistributions}) leaves little
doubt that the background concentration at $z=0.638$ ($\Delta z \sim
0.04$) corresponds to the matched-filter detection, with cluster
galaxies being distributed over nearly the entire field. This
conclusion is also supported by the image of the field shown in
Fig.~\ref{fig:img0048-2942}. The velocity dispersion of this system is
estimated to be $\sigma_v=1080\mathrm{km/s}$ indicating a massive
system.

Combined, the photometric (color and projected distribution) and
spectroscopic results provide strong evidence that we have detected a
real galaxy cluster at a redshift in excellent agreement with that
estimated by the matched-filter algorithm.

\subsection{EISJ0050-2941}

\begin{figure*}
\begin{center}
\resizebox{0.6\textwidth}{!}{\includegraphics{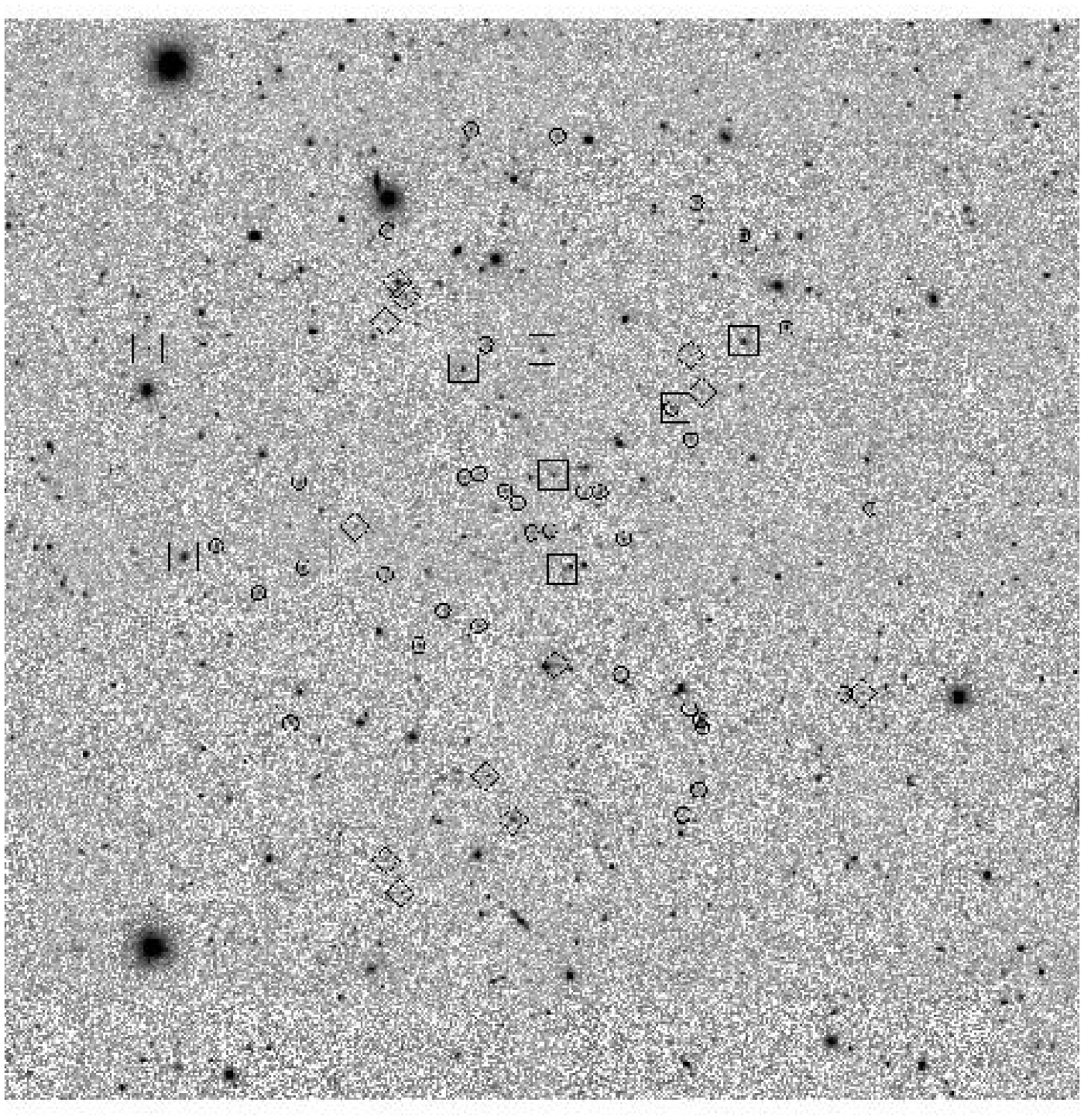}}
\caption{A 10$\times$10 arcmin cutout centered on the matched-filter
position of EISJ0050-2941. Symbols follow those in
Fig.~\protect\ref{fig:img0046-2951}. North is up, east to the left.}
\label{fig:img0050-2941}
\end{center}
\end{figure*}

In this field 55 redshifts have been measured and are shown in the
last panel in the first row of Fig.~\ref{fig:zdistributions}. The
distribution is considerably more complex than in the previous case,
with no single dominant peak discernible. The gap-technique identifies
6 groups with at least 3 members in the field, out of which two, with
comparable number of members, satisfy our significance criterion.
Information about these two systems is given in Table~\ref{tab:groups}
- a foreground system with 12 members at $z=0.558$ and a more distant
background system with 8 members at $z=0.616$. Note that both systems
are at considerably smaller redshifts than that estimated by the
matched filter and their projected spatial distribution (see
Fig.~\ref{fig:img0050-2941}) is scattered over the entire field. Taken
together this casts some doubt on the association of these density
enhancements in redshift space with the matched-filter detection at
$z_{MF}=0.9$. In addition, the colors listed in
Table~\ref{tab:cl_targets} point to a more distant
concentration. These colors correspond to the reddest of two peaks
detected by the color-slicing analysis by \cite{olsen00}. The
identification of two color peaks indicate the presence of two
superposed systems of which one is possibly more distant than
indicated by the spectroscopic redshifts.

In order to investigate this point further, we visually examined the
available imaging data (Fig.~\ref{fig:img0050-2941}), finding an
apparent clustering of faint galaxies very close to the position of
the matched-filter detection but only two of these galaxies have a
measured redshift (with $z\sim0.39$ and $z\sim0.58$).  Still there are
many more faint galaxies in the concentration and it is conceivable
that a more distant system may still exist lying behind two widely
scattered foreground systems. It is unclear at the present time
whether the matched-filter detection corresponds to the combination of
the two spectroscopically identified systems, with the faint galaxies
leading to an overestimate of the matched-filter redshift, or whether
the original detection is caused by a background concentration without
measured redshifts and the two identified systems lying in the
foreground. Clearly, with the available data alone it is not possible
to resolve this ambiguity which must await additional spectroscopic
observations in the field.

Regardless of the interpretation, the present results strongly suggest
the presence of two systems with estimated velocity dispersions of
1375 and 970~km/s for the foreground and background systems,
respectively, but with redshift offsets about twice as large as the
estimated accuracy.

\subsection{EISJ2236-4017}

\begin{figure*}
\begin{center}
\resizebox{0.6\textwidth}{!}{\includegraphics{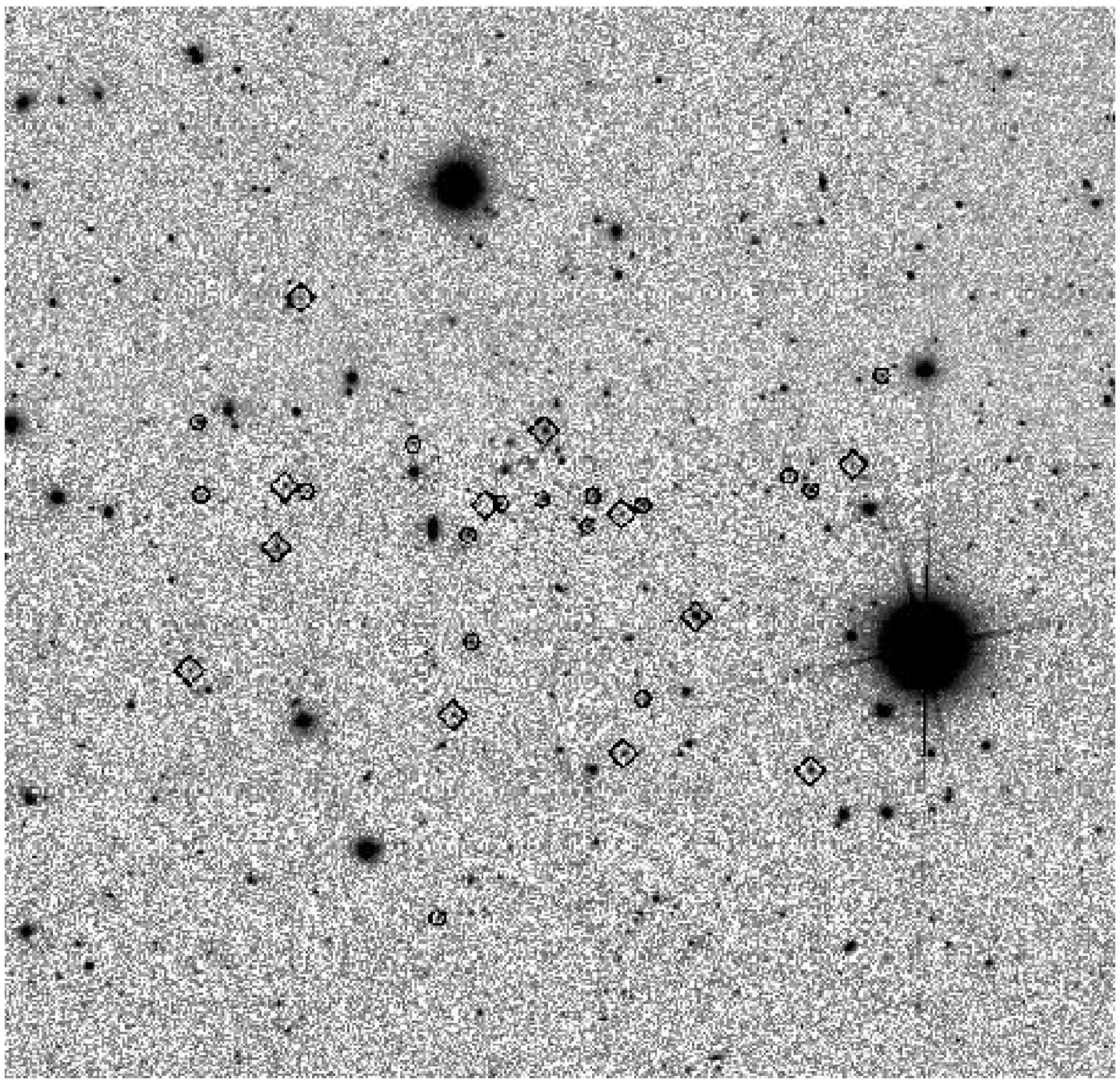}}
\caption{A 10$\times$10 arcmin cutout centered on the
matched-filter position of EISJ2236-4017. Symbols follow those in
Fig.~\protect\ref{fig:img0046-2951}, except that in this case only one
significant groups is found. North is up, east to the left.}
\label{fig:img2236-4017}
\end{center}
\end{figure*}

In this field 28 redshifts were measured and are shown in the left
panel in the second row of Fig.~\ref{fig:zdistributions}. As in the
case of EISJ0048-2942 the redshift distribution shows a distinct peak
at $z\sim0.5$ as well as the suggestion of other enhancements at
somewhat larger redshifts (\eg $z\sim 0.65$). In fact, the
gap-technique identifies 3 groups with at least 3 members in the
field. However, only one satisfies our significance criterion. The
properties of this system are given in Table~\ref{tab:groups}. It has
12 members and a redshift of 0.509 which, even though slightly
smaller, is consistent with that estimated by both the matched-filter
and the color-slicing technique. A cutout of the cluster region with
field and member galaxies marked is shown in
Fig.~\ref{fig:img2236-4017}. The system does not appear to be very
concentrated but has member galaxies being uniformly distributed, even
though the color slicing showed a strong peak at the position of the
matched-filter detection \citep{olsen00}. However, for this field not
all the masks were observed. In fact, the missing mask was the most
likely to include the cluster brightest members in the central
regions. This not only explains the significantly smaller number of
galaxies with measured redshift, but the lack of visible clustering
may also be due to the poor sampling achieved in this field. As can be
seen from the image cutout the region around the matched-filter
position at the center of the image is almost devoid of measured
redshifts. It is thus likely that the concentration of galaxies in the
center of the field corresponds to the matched-filter detection but
does not have any measured redshifts.

Regardless of the match with the matched-filter detection there is
evidence for the presence of a galaxy system at $z=0.509$ with a
velocity dispersion of 900~km/s.

\subsection{EISJ2249-3958}

\begin{figure*}
\begin{center}
\resizebox{0.6\textwidth}{!}{\includegraphics{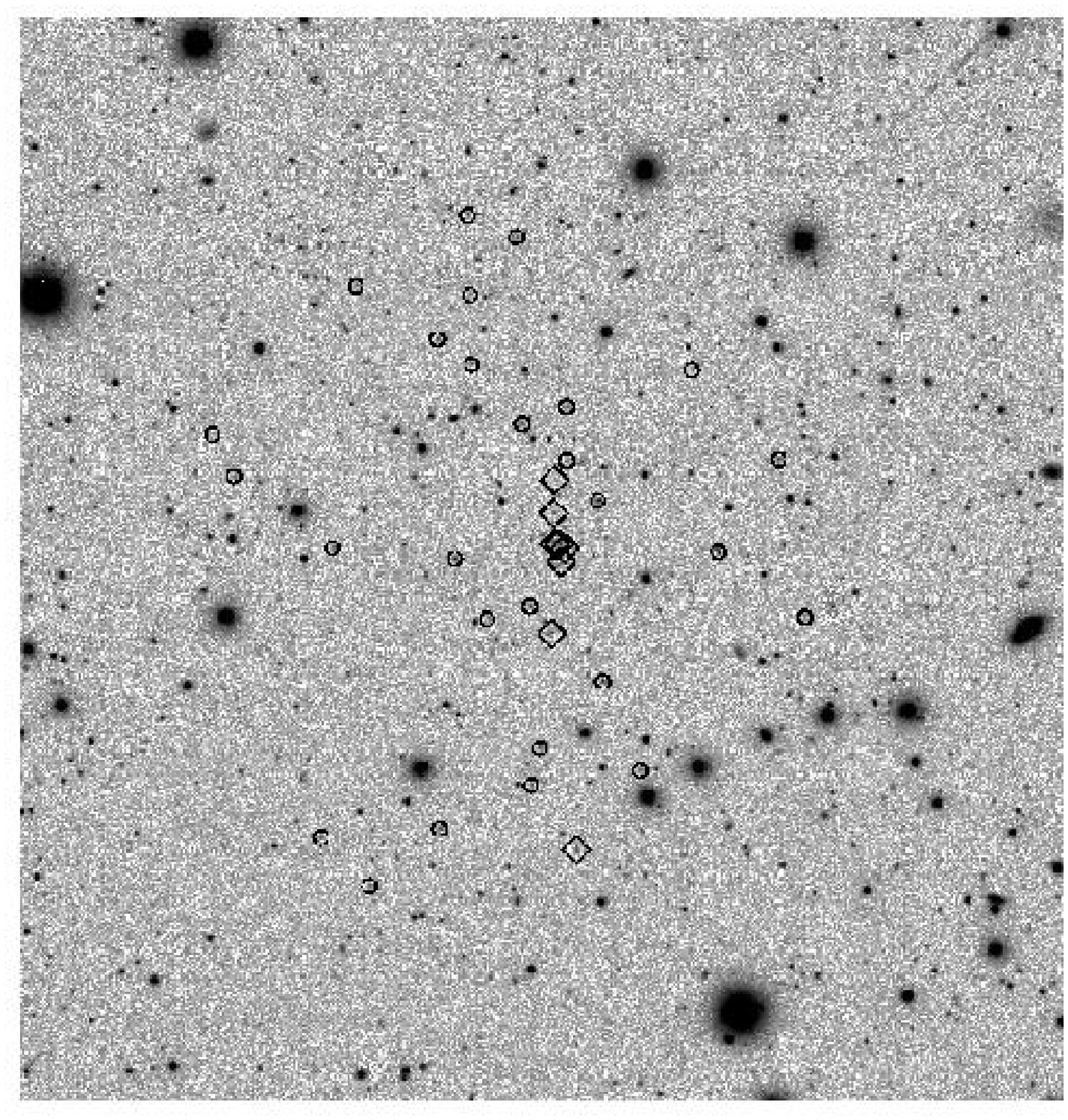}}
\caption{A 10$\times$10 arcmin cutout centered on the
matched-filter position of EISJ2249-3958. Symbols follow those in
Fig.~\protect\ref{fig:img2236-4017}. North is up, east to the left.}
\label{fig:img2249-3958}
\end{center}
\end{figure*}

In this field 35 redshifts were measured and are shown in the last row
of Fig.~\ref{fig:zdistributions}. The distribution shows a distinct
peak at $z\sim0.7$ as well as other smaller peaks both in the
foreground and background. Using the gap-technique, we indeed identify
five groups with at least 3 members, but only one is significant
according to the criteria adopted in this paper ($\sigma
\geq99\%$). As listed in Table~\ref{tab:groups} this cluster has a
redshift of $z=0.710$ ($\Delta z \sim -0.19$), somewhat smaller than
that estimated by the matched filter, and a velocity dispersion of
about 380~km/s, typical of groups. Inspection of the cone diagrams and
the $I$-band image (Fig.~\ref{fig:img2249-3958}) shows that seven out
of eight confirmed member galaxies lie along an elongated structure
extending only 2~ar\-cmin. The remaining galaxy is positioned along
the same axis but 2~arc\-min away from the rest. The concentrated
galaxies are the brightest ones found at the matched-filter position,
and thus likely to correspond to the original detection. The small
values of the velocity dispersion may be due to poor sampling or
alternatively that this density enhancement is associated to a
filament or a non-virialized cluster, instead of part of a relaxed
system. However, deciding among these various possibilities must
await further spectroscopic observations. For the time being, we
consider the detected system in redshift space to correspond to the
matched-filter detection of the projected distribution, which led to
an overestimate of the redshift.

\section{Discussion}
\label{sec:discussion}

The main objective of the present paper has been to extend the earlier
work of \cite{benoist02} and present the results of a spectroscopic
survey conducted at the VLT of the fields of 8 EIS candidate clusters
with redshifts $z\geq0.6$.  From the above analysis we find that in
three of the 5 fields considered here we identify two statistically
significant density enhancements and one in each of the two remaining
fields.  More importantly, the measured redshifts for these systems
range from $ 0.4~\lesssim~z~\lesssim~0.7$ and nearly all with velocity
dispersions typical of rich systems.  A less obvious question is
whether these detections are associated with the original
matched-filter detection. In general, one would say yes but in at
least one (but probably two) case(s) it appears that we have detected
a foreground system and that detected by the matched-filter technique
still needs to be confirmed by additional observations of fainter
galaxies. Another point that must await further observations is the
nature of the systems - namely, whether they form relaxed clusters, or
are part of proto-clusters before infall, or density enhancements
associated with filaments and walls. A preliminary effort in answering
these questions is presented by J{\o}rgensen et al. (2005, in
preparation).

Combining the present results with those compiled by \cite{benoist02},
our group has now studied the fields of 8 high-z candidate clusters,
with all leading to at least one confirmed system.  The properties of
all the detected systems are listed in
Table~\ref{tab:conf_EIS_cl}. The table gives: in Col.~1 the cluster field
name; in Col.~2 the matched-filter estimated redshift, $z_{MF}$,
whenever we believe that there is a match between the detection in
redshift and projected space; in Col.~3 the spectroscopic redshift,
$z_{spec}$, of the systems detected in redshift space; and in Col.~4
the estimated velocity dispersion of the system, $\sigma_v$.

For the six systems for which we believe to have identified the
counterpart of the matched-filter detection we find that the
difference between spectroscopic and estimated matched-filter
redshifts range from $\Delta z=z_{spec}-z_{MF}=-0.229$ to $\Delta z =
0.208$, with a mean offset of $\Delta z=-0.022$ and a standard
deviation of $\sim0.15$, therefore in excellent agreement with what
would be expected from the estimated errors of the algorithm.  This
result is valid all the way to the highest redshifts found in the
catalog and thus makes the EIS cluster candidate catalog a good source
for drawing high-z clusters for more detailed studies.

\begin{table}
\caption{Summary of confirmed EIS clusters from this work and from
\cite{benoist02}.}
\label{tab:conf_EIS_cl}
\begin{minipage}{\columnwidth}
\begin{center}
\begin{tabular}{lrrr}
\hline\hline
Cluster & $z_{MF}$ & $z_{spec}$ & $\sigma_v \mathrm{(km/s)}$\\
\hline
EISJ0046-2930$^*$ & 0.6 & 0.808 & 1170\\ 
EISJ0046-2951 &     & 0.614 & 1400 \\ 
              & 0.9 & 0.671 & 865  \\ 
EISJ0048-2942 &     & 0.402 & 1000 \\ 
              & 0.6 & 0.638 & 1080 \\ 
EISJ0050-2941 &     & 0.559 & 1375 \\
              &     & 0.617 & 970 \\
EISJ0533-2412$^*$ & 1.3 & 1.301 & $-$\\ 
EISJ0954-2023$^*$ &     & 0.948 & 200\\
              & 1.1 & 1.141 & 290\\ 
EISJ2236-4017 &     & 0.509 & 900\\
EISJ2249-3958 & 0.9 & 0.710 & 380 \\ 
\hline
\end{tabular}
\end{center}
* The spectroscopic confirmations of these systems were reported in
\cite{benoist02}.
\end{minipage}
\end{table}

\section {Summary}
\label{sec:summary}

In this paper we have presented the results of spectroscopic
observations conducted with FORS1 at the VLT in the fields of 5 high-z
($z\geq0.6$) cluster candidates identified by applying the
matched-filter algorithm to the images from the EIS-WIDE $I$-band
survey. The presence of galaxy clusters were supported by a
color-slicing analysis targeting the individual detections. We find at
least one significant system in all fields with redshifts in the range
$0.40 < z < 0.71$ and from 8 to 33 confirmed cluster members. All
systems, except one, have velocity dispersions $\gtrsim
800\mathrm{km/s}$ typical of that of rich clusters.  Despite the
intrinsic ambiguity of uniquely associating a significant density
enhancement in redshift space with a detection in the projected
distribution, the agreement of the matched filter and spectroscopic
redshifts is, on average, excellent even though the matched filter has
a tendency to overestimate them at higher redshifts.  The results of
this paper together with others of this series, strongly suggest that
nearly all of the EIS candidate clusters identified applying the
matched-filter algorithm to the $I$-band galaxy catalogs are
associated with real density enhancements in redshift space regardless
of the redshift domain. We conclude that the EIS Cluster Candidate
Catalog is an excellent starting point to build a statistical sample
of galaxy clusters at different redshifts for further investigation,
complementing in many ways samples based on X-ray selection.

\void{
 In this paper we have presented the results of spectrocospic
observations conducted with FORS at the VLT in the fields of 5 high-z
($z>0.6$) cluster candidates identified by applying the matched-filter
algorithm to the images from the EIS-WIDE $I$-band survey.  Despite of
the intrinsic ambiguity of uniquely associating a significant density
enhancement in redshift space with a detection in the projected
distribution, we find at least one significant system in all fields
with redshifts in the range $0.5 < z < 0.71$ and from 8 to 33
confirmed cluster members. All systems, except one, have velocity
dispersions $\gtrsim 900\mathrm{km/s}$ typical of that of rich
clusters. The agreement of the matched-filter and spectroscopic
redshifts is excellent for systems at z=0.6, but the matched filter
seem to over estimate them at higher redshifts.  The results of this
paper together with others of this series, strongly suggests that
nearly all of the EIS candidate clusters identified using the
matched-filter analysis to the $I$-band images are associated with
real density enhancements in redshift space regardless of the redshift
domain, being therefore an excellent starting point to build a
statistical sample of cluster of galaxies at different redshifts for
further investigation, complementing in many ways samples based on
X-ray selection.
 }


\begin{acknowledgements}
 We would like to thank the referee for many useful comments, which
 helped improve the manuscript.  LFO acknowledges financial support
 from the Carlsberg Foundation, the Danish Natural Sciences Research
 Council and the Poincar\'e fellowship program at Observatoire de la
 C\^ote d'Azur.
\end{acknowledgements}

\bibliographystyle{../../../aa}
\bibliography{/home/lisbeth/tex/lisbeth_ref}

\newpage

\begin{appendix}

\section{Measured redshifts}
\label{app:gal_redshifts}

\begin{table}
\caption{The list of measured redshifts for the cluster EISJ0046-2951.}
\label{tab:EIS0046-2951}
\begin{tabular}{rrrrr}
\hline\hline
ID & $\alpha (J2000)$ & $\delta (J2000)$ & $I$ & $z$\\
\hline
      1 & 00:45:42.318 & -29:48:14.68 &   20.52 &   0.8206 \\
      2 & 00:46:00.877 & -29:49:27.56 &   21.93 &   0.6099 \\
      3 & 00:45:57.779 & -29:49:30.50 &   21.51 &   0.6650 \\
      4 & 00:45:48.689 & -29:49:39.00 &   21.47 &   0.6255 \\
      5 & 00:45:53.937 & -29:49:51.54 &   19.92 &   0.6088 \\
      6 & 00:45:59.547 & -29:50:30.09 &   21.22 &   0.7832 \\
      7 & 00:46:01.161 & -29:51:35.91 &   21.46 &   0.3099 \\
      8 & 00:46:00.264 & -29:52:29.38 &   18.59 &   0.2805 \\
      9 & 00:45:58.466 & -29:52:40.12 &   21.10 &   0.8205 \\
     10 & 00:46:02.725 & -29:53:55.33 &   21.46 &   0.7310 \\
     11 & 00:46:02.932 & -29:53:55.28 &   22.60 &   0.7310 \\
     12 & 00:46:00.057 & -29:53:58.11 &   21.40 &   0.7612 \\
     13 & 00:45:47.826 & -29:47:38.91 &   20.67 &   0.2171 \\
     14 & 00:46:12.736 & -29:47:41.42 &   21.92 &   0.6113 \\
     15 & 00:46:15.350 & -29:47:43.35 &   21.77 &   0.6805 \\
     16 & 00:45:48.226 & -29:47:54.55 &   20.89 &   0.6288 \\
     17 & 00:45:50.702 & -29:48:02.36 &   20.19 &   0.4410 \\
     18 & 00:45:43.652 & -29:48:14.77 &   20.69 &   0.8229 \\
     19 & 00:45:53.812 & -29:48:13.45 &   21.17 &   0.4072 \\
     20 & 00:45:59.666 & -29:48:19.38 &   21.66 &   0.4912 \\
     21 & 00:46:13.850 & -29:48:19.92 &   21.22 &   0.2151 \\
     22 & 00:46:14.864 & -29:48:24.76 &   20.24 &   0.5618 \\
     23 & 00:46:11.498 & -29:48:24.65 &   20.74 &   0.1922 \\
     24 & 00:46:03.188 & -29:49:11.56 &   18.20 &   0.6130 \\
     25 & 00:46:03.522 & -29:49:08.03 &   19.34 &   0.6171 \\
     26 & 00:45:54.351 & -29:48:51.43 &   20.75 &   0.6130 \\
     27 & 00:45:48.286 & -29:48:51.89 &   21.37 &   0.2190 \\
     28 & 00:45:59.957 & -29:49:02.26 &   20.78 &   0.6135 \\
     29 & 00:45:55.851 & -29:49:06.60 &   20.61 &   0.5147 \\
     30 & 00:45:54.989 & -29:49:04.22 &   20.60 &   0.5137 \\
     31 & 00:46:03.939 & -29:49:08.80 &   20.06 &   0.1322 \\
     32 & 00:46:10.209 & -29:49:06.74 &   21.19 &   0.6071 \\
     33 & 00:46:14.783 & -29:49:08.51 &   20.97 &   0.5636 \\
     34 & 00:46:08.292 & -29:49:13.98 &   19.58 &   0.6092 \\
     35 & 00:46:05.373 & -29:49:16.20 &   21.74 &   0.6232 \\
     36 & 00:46:09.963 & -29:49:14.78 &   21.88 &   0.3194 \\
     37 & 00:46:05.730 & -29:49:16.89 &   20.93 &   0.1461 \\
     38 & 00:46:08.955 & -29:49:18.13 &   21.22 &   0.6132 \\
     39 & 00:46:18.326 & -29:49:24.38 &   21.95 &   0.3853 \\
     40 & 00:46:07.438 & -29:49:35.66 &   21.62 &   0.4080 \\
     41 & 00:46:14.058 & -29:49:38.96 &   21.48 &   0.6140 \\
     42 & 00:46:06.995 & -29:49:42.90 &   22.07 &   0.7433 \\
     43 & 00:46:06.295 & -29:49:54.80 &   20.10 &   0.6717 \\
     44 & 00:46:06.629 & -29:49:54.86 &   20.74 &   0.5117 \\
     45 & 00:46:03.327 & -29:50:12.70 &   19.71 &   0.4427 \\
     46 & 00:46:08.951 & -29:50:15.14 &   21.73 &   0.8789 \\
     47 & 00:46:12.814 & -29:50:33.70 &   20.51 &   0.6108 \\
     48 & 00:46:06.219 & -29:50:39.72 &   21.72 &   0.5584 \\
     49 & 00:46:10.870 & -29:51:06.54 &   19.96 &   0.7653 \\
     50 & 00:46:05.844 & -29:51:07.94 &   19.94 &   0.5402 \\
     51 & 00:46:08.049 & -29:51:09.97 &   21.85 &   0.6650 \\
     52 & 00:46:14.610 & -29:51:13.81 &   20.48 &   0.6270 \\
     53 & 00:46:07.916 & -29:51:24.57 &   19.75 &   0.6735 \\
     54 & 00:46:09.105 & -29:51:25.28 &   21.86 &   0.6760 \\
     55 & 00:46:06.317 & -29:51:27.77 &   21.02 &   0.6684 \\
     56 & 00:46:08.982 & -29:51:30.55 &   21.92 &   0.3149 \\
     57 & 00:46:08.418 & -29:51:37.73 &   20.59 &   0.2166 \\
     58 & 00:46:04.555 & -29:52:53.72 &   21.04 &   0.5379 \\
     59 & 00:46:07.854 & -29:53:22.11 &   21.93 &   0.6671 \\
     60 & 00:46:03.484 & -29:53:24.55 &   22.01 &   0.2803 \\
\hline
\end{tabular}
\end{table}

\addtocounter{table}{-1}

\begin{table}
\caption{\it -- Continued}
\begin{tabular}{rrrrr}
\hline\hline
ID & $\alpha (J2000)$ & $\delta (J2000)$ & $I$ & $z$\\
\hline
     61 & 00:46:06.670 & -29:53:24.75 &   21.20 &   0.5491 \\
     62 & 00:46:11.473 & -29:51:40.39 &   20.95 &   0.6263 \\
     63 & 00:46:08.000 & -29:51:41.17 &   21.33 &   0.6725 \\
     64 & 00:46:06.194 & -29:51:36.85 &   22.28 &   0.4658 \\
     65 & 00:46:06.225 & -29:51:55.60 &   21.29 &   0.7318 \\
     66 & 00:46:04.053 & -29:51:50.77 &   21.83 &   0.7313 \\
     67 & 00:46:13.244 & -29:51:50.10 &   21.29 &   0.7078 \\
     68 & 00:46:04.614 & -29:52:16.66 &   21.81 &   1.1291 \\
     69 & 00:46:03.632 & -29:52:20.23 &   21.72 &   0.7393 \\
     70 & 00:46:04.778 & -29:52:44.59 &   22.00 &   0.3329 \\
     71 & 00:46:10.045 & -29:53:09.80 &   21.09 &   0.6704 \\
\hline
\end{tabular}
\end{table}

\begin{table}
\caption{The list of measured redshifts for the cluster EISJ0048-2942.}
\label{tab:EIS0048-2942}
\begin{tabular}{rrrrr}
\hline\hline
ID & $\alpha (J2000)$ & $\delta (J2000)$ & $I$ & $z$\\
\hline
      1 & 00:48:30.904 & -29:45:05.18 &   20.63 &   0.6400 \\
      2 & 00:48:31.854 & -29:45:12.25 &   20.99 &   0.6307 \\
      3 & 00:48:28.238 & -29:45:31.60 &   21.21 &   0.3765 \\
      4 & 00:48:22.606 & -29:39:00.96 &   18.94 &   0.2015 \\
      5 & 00:48:21.332 & -29:41:02.73 &   21.93 &   0.6447 \\
      6 & 00:48:20.495 & -29:42:14.72 &   21.19 &   0.6350 \\
      7 & 00:48:22.138 & -29:42:23.62 &   21.27 &   0.6973 \\
      8 & 00:48:20.672 & -29:42:44.22 &   19.80 &   0.6086 \\
      9 & 00:48:24.625 & -29:42:55.23 &   21.39 &   0.3528 \\
     10 & 00:48:18.750 & -29:43:16.91 &   20.20 &   0.5421 \\
     11 & 00:48:19.632 & -29:43:55.35 &   19.62 &   0.5398 \\
     12 & 00:48:21.985 & -29:44:17.94 &   20.50 &   0.5418 \\
     13 & 00:48:20.124 & -29:44:37.72 &   21.20 &   0.6252 \\
     14 & 00:48:37.912 & -29:41:15.66 &   22.01 &   0.6426 \\
     15 & 00:48:31.682 & -29:41:25.85 &   22.28 &   0.1319 \\
     16 & 00:48:26.160 & -29:41:25.79 &   21.71 &   0.2119 \\
     17 & 00:48:29.152 & -29:41:30.69 &   21.99 &   0.3293 \\
     18 & 00:48:32.630 & -29:41:31.61 &   21.96 &   0.6508 \\
     19 & 00:48:40.353 & -29:41:36.68 &   20.70 &   8.8888 \\
     20 & 00:48:29.387 & -29:41:34.08 &   21.73 &   0.2251 \\
     21 & 00:48:35.571 & -29:41:37.44 &   20.67 &   0.6355 \\
     22 & 00:48:34.576 & -29:41:36.23 &   21.30 &   0.6332 \\
     23 & 00:48:39.006 & -29:41:47.05 &   21.68 &   0.6412 \\
     24 & 00:48:41.578 & -29:41:51.26 &   20.68 &   0.4411 \\
     25 & 00:48:26.023 & -29:41:53.37 &   21.20 &   0.6994 \\
     26 & 00:48:37.715 & -29:41:53.85 &   20.86 &   0.6383 \\
     27 & 00:48:41.216 & -29:41:59.95 &   20.50 &   0.3989 \\
     28 & 00:48:37.907 & -29:41:59.43 &   21.81 &   0.5175 \\
     29 & 00:48:32.699 & -29:42:04.09 &   21.13 &   0.6375 \\
     30 & 00:48:42.141 & -29:42:02.41 &   20.66 &   0.5476 \\
     31 & 00:48:31.883 & -29:42:00.38 &   20.97 &   0.6393 \\
     32 & 00:48:29.181 & -29:42:00.67 &   21.85 &   0.6390 \\
     33 & 00:48:31.385 & -29:42:10.05 &   19.31 &   0.6409 \\
     34 & 00:48:29.308 & -29:42:08.18 &   21.72 &   0.5476 \\
     35 & 00:48:30.050 & -29:42:11.05 &   21.09 &   0.6495 \\
     36 & 00:48:42.060 & -29:42:14.79 &   21.48 &   0.6351 \\
     37 & 00:48:27.635 & -29:42:15.38 &   21.44 &   0.3962 \\
     38 & 00:48:39.379 & -29:42:27.08 &   19.92 &   0.4600 \\
     39 & 00:48:29.671 & -29:42:26.64 &   21.52 &   0.4363 \\
     40 & 00:48:35.710 & -29:42:31.90 &   21.63 &   0.7419 \\
     41 & 00:48:34.754 & -29:42:38.04 &   21.77 &   0.1975 \\
     42 & 00:48:35.763 & -29:42:43.93 &   21.06 &   0.6506 \\
     43 & 00:48:40.461 & -29:42:51.47 &   21.42 &   0.5476 \\
     44 & 00:48:30.271 & -29:42:58.45 &   22.03 &   0.3786 \\
     45 & 00:48:43.648 & -29:43:03.87 &   19.62 &   0.7632 \\
     46 & 00:48:33.060 & -29:43:07.41 &   20.65 &   0.6319 \\
     47 & 00:48:38.103 & -29:43:08.39 &   20.84 &   0.6379 \\
     48 & 00:48:30.788 & -29:43:16.62 &   19.98 &   0.6357 \\
     49 & 00:48:40.006 & -29:43:16.93 &   21.19 &   0.6369 \\
     50 & 00:48:42.113 & -29:43:20.91 &   21.58 &   0.7665 \\
     51 & 00:48:41.507 & -29:43:23.86 &   20.63 &   0.6987 \\
     52 & 00:48:37.402 & -29:43:41.01 &   21.41 &   0.6319 \\
     53 & 00:48:41.557 & -29:43:42.44 &   21.03 &   8.8888 \\
     54 & 00:48:37.266 & -29:43:44.39 &   20.56 &   0.6370 \\
     55 & 00:48:26.997 & -29:43:43.69 &   22.13 &   0.6790 \\
     56 & 00:48:31.317 & -29:43:45.36 &   21.20 &   0.6275 \\
     57 & 00:48:39.264 & -29:43:53.00 &   20.34 &   0.6423 \\
     58 & 00:48:38.473 & -29:43:54.98 &   20.37 &   0.6754 \\
     59 & 00:48:37.685 & -29:44:00.39 &   20.84 &   0.6352 \\
     60 & 00:48:35.311 & -29:44:17.78 &   20.82 &   0.6364 \\
\hline
\end{tabular}
\end{table}

\addtocounter{table}{-1}

\begin{table}
\caption{\it -- Continued}
\begin{tabular}{rrrrr}
\hline\hline
ID & $\alpha (J2000)$ & $\delta (J2000)$ & $I$ & $z$\\
\hline
     61 & 00:48:31.112 & -29:44:19.50 &   21.68 &   0.8626 \\
     62 & 00:48:40.986 & -29:44:24.89 &   21.95 &   0.3015 \\
     63 & 00:48:33.184 & -29:44:24.90 &   21.08 &   0.5002 \\
     64 & 00:48:28.204 & -29:44:39.79 &   20.24 &   0.4082 \\
     65 & 00:48:31.408 & -29:44:46.13 &   22.01 &   0.6301 \\
     66 & 00:48:30.987 & -29:44:59.40 &   21.30 &   0.8240 \\
     67 & 00:48:30.116 & -29:39:06.04 &   20.13 &   0.4043 \\
     68 & 00:48:27.736 & -29:39:17.74 &   21.62 &   8.8888 \\
     69 & 00:48:33.042 & -29:39:37.73 &   21.22 &   0.5217 \\
     70 & 00:48:39.234 & -29:39:59.12 &   20.21 &   0.4011 \\
     71 & 00:48:41.374 & -29:40:10.57 &   21.74 &   0.4743 \\
     72 & 00:48:27.598 & -29:40:23.29 &   20.71 &   0.6379 \\
     73 & 00:48:40.952 & -29:40:27.39 &   20.55 &   0.3967 \\
     74 & 00:48:44.138 & -29:40:45.69 &   20.00 &   0.5312 \\
     75 & 00:48:36.816 & -29:40:47.07 &   22.22 &   0.6350 \\
     76 & 00:48:36.927 & -29:40:52.09 &   20.94 &   0.6328 \\
     77 & 00:48:29.378 & -29:40:51.35 &   20.70 &   0.6450 \\
     78 & 00:48:48.161 & -29:42:50.69 &   20.75 &   0.4071 \\
     79 & 00:48:46.908 & -29:43:08.91 &   20.51 &   0.4604 \\
     80 & 00:48:47.426 & -29:43:08.96 &   20.79 &   0.2197 \\
\hline
\end{tabular}
\end{table}

\begin{table}
\caption{The list of measured redshifts for the cluster EISJ0050-2941.}
\label{tab:EIS0050-2941}
\begin{tabular}{rrrrr}
\hline\hline
ID & $\alpha (J2000)$ & $\delta (J2000)$ & $I$ & $z$\\
\hline
      1 & 00:50:03.983 & -29:41:19.24 &   21.54 &   0.3956 \\
      2 & 00:50:04.744 & -29:41:20.27 &   21.67 &   0.5811 \\
      3 & 00:50:00.828 & -29:41:24.02 &   20.74 &   0.8226 \\
      4 & 00:50:03.389 & -29:41:40.52 &   21.97 &   0.6176 \\
      5 & 00:50:03.762 & -29:42:33.94 &   21.00 &   0.5690 \\
      6 & 00:50:00.936 & -29:42:39.07 &   21.98 &   0.5868 \\
      7 & 00:49:50.655 & -29:42:49.86 &   22.00 &   0.5690 \\
      8 & 00:49:51.353 & -29:42:50.19 &   21.43 &   0.5007 \\
      9 & 00:49:57.570 & -29:43:04.84 &   19.36 &   0.5267 \\
     10 & 00:49:58.044 & -29:42:58.77 &   21.81 &   0.1647 \\
     11 & 00:49:57.468 & -29:43:08.72 &   21.43 &   0.5265 \\
     12 & 00:49:57.644 & -29:43:43.54 &   22.04 &   0.5233 \\
     13 & 00:49:58.338 & -29:43:57.60 &   21.12 &   0.5077 \\
     14 & 00:50:03.611 & -29:37:39.81 &   21.48 &   0.3008 \\
     15 & 00:50:07.277 & -29:37:36.06 &   21.31 &   0.3128 \\
     16 & 00:49:57.687 & -29:38:17.10 &   20.65 &   0.5806 \\
     17 & 00:49:55.689 & -29:38:35.31 &   19.98 &   0.5102 \\
     18 & 00:50:10.860 & -29:38:32.40 &   21.06 &   0.4358 \\
     19 & 00:50:10.442 & -29:39:00.98 &   20.17 &   0.5578 \\
     20 & 00:50:10.093 & -29:39:08.55 &   19.96 &   0.5574 \\
     21 & 00:50:10.950 & -29:39:22.25 &   22.10 &   0.5540 \\
     22 & 00:49:53.856 & -29:39:26.65 &   20.18 &   0.7331 \\
     23 & 00:49:55.687 & -29:39:34.21 &   20.01 &   0.6281 \\
     24 & 00:50:06.658 & -29:39:35.57 &   21.38 &   0.7028 \\
     25 & 00:50:21.071 & -29:39:37.27 &   21.21 &   0.6159 \\
     26 & 00:50:04.286 & -29:39:39.02 &   21.27 &   0.6128 \\
     27 & 00:49:57.957 & -29:39:41.95 &   21.30 &   0.5506 \\
     28 & 00:50:07.671 & -29:39:48.92 &   20.37 &   0.6132 \\
     29 & 00:49:58.758 & -29:40:11.11 &   21.70 &   0.6564 \\
     30 & 00:49:57.452 & -29:40:02.42 &   21.36 &   0.5506 \\
     31 & 00:49:58.578 & -29:40:10.92 &   22.71 &   0.6234 \\
     32 & 00:49:57.959 & -29:40:28.87 &   20.57 &   0.5948 \\
     33 & 00:50:03.780 & -29:40:48.08 &   20.45 &   0.6149 \\
     34 & 00:50:06.977 & -29:40:47.66 &   21.50 &   0.5980 \\
     35 & 00:50:14.650 & -29:40:52.39 &   20.69 &   0.6996 \\
     36 & 00:50:07.623 & -29:40:49.35 &   21.92 &   0.4943 \\
     37 & 00:50:05.888 & -29:40:56.94 &   21.80 &   0.2179 \\
     38 & 00:50:01.805 & -29:40:57.71 &   20.97 &   0.3356 \\
     39 & 00:50:02.487 & -29:40:58.06 &   22.01 &   0.3359 \\
     40 & 00:50:05.322 & -29:41:03.77 &   22.04 &   0.3137 \\
     41 & 00:49:50.259 & -29:41:07.57 &   21.76 &   0.4414 \\
     42 & 00:50:12.261 & -29:41:16.64 &   22.07 &   0.5468 \\
     43 & 00:50:18.204 & -29:41:26.93 &   20.83 &   0.5315 \\
     44 & 00:50:19.585 & -29:41:33.25 &   19.11 &   0.6120 \\
     45 & 00:50:14.502 & -29:41:39.32 &   20.70 &   0.5304 \\
     46 & 00:50:10.993 & -29:41:43.32 &   21.25 &   0.7689 \\
     47 & 00:50:16.403 & -29:41:53.57 &   21.46 &   0.4410 \\
     48 & 00:50:08.547 & -29:42:03.15 &   21.62 &   0.4962 \\
     49 & 00:50:06.986 & -29:42:11.97 &   21.07 &   0.5995 \\
     50 & 00:50:09.569 & -29:42:22.78 &   20.41 &   0.3379 \\
     51 & 00:50:15.047 & -29:43:05.89 &   20.29 &   0.3156 \\
     52 & 00:50:06.715 & -29:43:36.08 &   20.84 &   0.5629 \\
     53 & 00:50:05.556 & -29:44:00.26 &   20.18 &   0.5621 \\
     54 & 00:50:11.011 & -29:44:22.21 &   21.17 &   0.5575 \\
     55 & 00:50:10.379 & -29:44:40.59 &   21.12 &   0.5644 \\
\hline
\end{tabular}
\end{table}

\begin{table}
\caption{The list of measured redshifts for the cluster EISJ2236-4017.}
\label{tab:EIS2236-4017}
\begin{tabular}{rrrrr}
\hline\hline
ID & $\alpha (J2000)$ & $\delta (J2000)$ & $I$ & $z$\\
\hline
      1 & 22:36:23.524 & -40:21:22.82 &   20.23 &   0.5727 \\
      2 & 22:36:02.714 & -40:16:20.02 &   20.61 &   0.6503 \\
      3 & 22:36:18.851 & -40:16:53.88 &   18.86 &   0.5062 \\
      4 & 22:36:04.007 & -40:17:09.27 &   20.74 &   0.4988 \\
      5 & 22:36:07.049 & -40:17:16.12 &   20.40 &   0.6250 \\
      6 & 22:36:06.005 & -40:17:23.87 &   19.14 &   0.3308 \\
      7 & 22:36:16.462 & -40:17:29.03 &   19.86 &   0.6248 \\
      8 & 22:36:14.059 & -40:17:33.63 &   19.21 &   0.5969 \\
      9 & 22:36:18.895 & -40:17:31.54 &   19.75 &   0.4471 \\
     10 & 22:36:20.863 & -40:17:34.28 &   20.63 &   0.5571 \\
     11 & 22:36:21.606 & -40:17:35.22 &   21.30 &   0.5089 \\
     12 & 22:36:15.069 & -40:17:38.70 &   20.23 &   0.5126 \\
     13 & 22:36:16.693 & -40:17:45.61 &   20.44 &   0.6455 \\
     14 & 22:36:22.419 & -40:17:51.96 &   19.24 &   0.6256 \\
     15 & 22:36:11.417 & -40:18:34.50 &   18.22 &   0.5139 \\
     16 & 22:36:22.142 & -40:18:50.34 &   19.66 &   0.0751 \\
     17 & 22:36:13.885 & -40:19:20.01 &   20.50 &   0.5680 \\
     18 & 22:36:22.978 & -40:19:31.11 &   20.26 &   0.5095 \\
     19 & 22:36:14.760 & -40:19:50.46 &   19.40 &   0.5143 \\
     20 & 22:36:05.780 & -40:19:57.94 &   19.93 &   0.5097 \\
     21 & 22:36:30.651 & -40:15:43.28 &   19.97 &   0.5143 \\
     22 & 22:36:35.441 & -40:16:52.38 &   19.82 &   0.2381 \\
     23 & 22:36:25.117 & -40:17:02.84 &   19.85 &   0.6462 \\
     24 & 22:36:31.255 & -40:17:27.34 &   20.61 &   0.5061 \\
     25 & 22:36:30.178 & -40:17:29.66 &   21.16 &   0.4262 \\
     26 & 22:36:35.239 & -40:17:32.28 &   20.47 &   0.2026 \\
     27 & 22:36:31.602 & -40:18:00.80 &   19.43 &   0.5039 \\
     28 & 22:36:35.651 & -40:19:09.21 &   19.91 &   0.5070 \\
\hline
\end{tabular}
\end{table}

\begin{table}
\caption{The list of measured redshifts for the cluster EISJ2249-3958.}
\label{tab:EIS2249-3958}
\begin{tabular}{rrrrr}
\hline\hline
ID & $\alpha (J2000)$ & $\delta (J2000)$ & $I$ & $z$\\
\hline
      1 & 22:49:33.164 & -39:59:56.62 &   21.70 &   0.3450 \\
      2 & 22:49:28.265 & -40:00:10.34 &   21.59 &   0.6721 \\
      3 & 22:49:33.640 & -40:00:17.34 &   22.12 &   0.5030 \\
      4 & 22:49:24.232 & -40:00:34.68 &   21.77 &   8.8888 \\
      5 & 22:49:38.161 & -40:00:40.94 &   21.55 &   0.2749 \\
      6 & 22:49:31.482 & -40:00:54.10 &   21.43 &   0.7061 \\
      7 & 22:49:25.362 & -39:55:42.70 &   21.22 &   8.8888 \\
      8 & 22:49:37.812 & -39:56:04.71 &   21.36 &   0.7471 \\
      9 & 22:49:24.025 & -39:56:11.05 &   21.57 &   8.8888 \\
     10 & 22:49:36.162 & -39:56:19.42 &   21.86 &   0.7245 \\
     11 & 22:49:25.378 & -39:56:24.84 &   21.82 &   0.5027 \\
     12 & 22:49:31.537 & -39:56:44.18 &   20.98 &   0.6568 \\
     13 & 22:49:33.746 & -39:56:53.37 &   21.77 &   0.4274 \\
     14 & 22:49:48.942 & -39:56:55.95 &   22.49 &   1.0059 \\
     15 & 22:49:21.179 & -39:57:16.22 &   21.29 &   0.5286 \\
     16 & 22:49:31.593 & -39:57:14.72 &   21.97 &   0.5681 \\
     17 & 22:49:47.923 & -39:57:19.96 &   21.49 &   0.7468 \\
     18 & 22:49:32.189 & -39:57:25.29 &   21.68 &   0.7103 \\
     19 & 22:49:30.078 & -39:57:37.36 &   20.29 &   0.5698 \\
     20 & 22:49:32.265 & -39:57:44.10 &   21.82 &   0.7108 \\
     21 & 22:49:32.247 & -39:58:01.47 &   19.80 &   0.7127 \\
     22 & 22:49:31.747 & -39:58:04.26 &   21.40 &   0.7088 \\
     23 & 22:49:32.028 & -39:58:01.61 &   20.83 &   0.7075 \\
     24 & 22:49:24.222 & -39:58:07.66 &   19.97 &   0.5260 \\
     25 & 22:49:43.154 & -39:58:01.43 &   22.10 &   0.7286 \\
     26 & 22:49:23.216 & -39:58:07.41 &   21.16 &   8.8888 \\
     27 & 22:49:31.932 & -39:58:11.74 &   20.53 &   0.7093 \\
     28 & 22:49:37.168 & -39:58:08.80 &   20.84 &   0.8453 \\
     29 & 22:49:33.538 & -39:58:36.52 &   21.11 &   0.5555 \\
     30 & 22:49:35.627 & -39:58:43.12 &   22.24 &   0.9299 \\
     31 & 22:49:20.042 & -39:58:45.46 &   21.65 &   0.5012 \\
     32 & 22:49:43.364 & -39:58:48.64 &   22.01 &   8.8888 \\
     33 & 22:49:32.475 & -39:58:52.08 &   21.06 &   0.7122 \\
     34 & 22:49:35.232 & -39:58:55.16 &   21.64 &   8.8888 \\
     35 & 22:49:30.020 & -39:59:19.77 &   20.82 &   0.5557 \\
     36 & 22:49:43.991 & -40:00:44.86 &   22.30 &   0.7470 \\
     37 & 22:49:41.654 & -40:01:12.55 &   21.97 &   0.9072 \\
     38 & 22:49:36.215 & -39:54:55.04 &   21.75 &   0.7277 \\
     39 & 22:49:33.834 & -39:55:07.99 &   21.69 &   0.4917 \\
     40 & 22:49:36.168 & -39:55:40.36 &   22.20 &   0.6173 \\
     41 & 22:49:41.750 & -39:55:33.83 &   22.25 &   0.5281 \\
     42 & 22:49:40.305 & -39:55:14.17 &   22.21 &   8.8888 \\
\hline
\end{tabular}
\end{table}

%

\end{appendix}

\end{document}